\input{epsf}

\documentstyle[amsfonts,amssymb,amstex,aps]{revtex}

\begin{document}

\title{On the Completeness of the Quasinormal Modes of the
 P\"{o}schl-Teller Potential}

\author{Horst R. Beyer \\
 Max Planck Institute for Gravitational Physics, \\
 Albert Einstein Institute, \\
 D-14473 Potsdam, Germany\\
Fax: (0)711/685-6282}


\maketitle

\begin{abstract}
 The completeness of the quasinormal modes of the wave equation with
 P\"{o}schl-Teller potential is investigated. A main result is that
 after a large enough time $t_0$, the solutions of this equation
 corresponding to $C^{\infty}$-data with compact support can be
 expanded uniformly in time with respect to the quasinormal modes,
 thereby leading to absolutely convergent series. Explicit estimates
 for $t_0$ depending on both the support of the data and the point of
 observation are given. For the particular case of an ``early'' time
 and zero distance between the support of the data and observational
 point, it is shown that the corresponding series is {\it not}
 absolutely convergent, and hence that there is no associated sum
 which is independent of the order of summation.
\end{abstract}


\section{General Introduction}

The description of a compact classical system often leads to the
consideration of a ``small'' perturbation of some special solution of
its evolution equations. Expanding around this solution leads to a
linear evolution equation for some perturbed quantities which
characterise the system. For a system with no explicit time
dependence, a further step consists of finding the {\it normal mode
 solutions} of the evolution equation satisfying certain physical
boundary conditions. To provide a complete description of the system
under small perturbations, every solution of the linear equation
satisfying the boundary conditions must have an expansion in terms of
these modes.

To my knowledge the only well-developed mathematical framework to date
for deciding such a ``completeness'' question is provided by the
spectral theory of linear operators in Hilbert spaces. This is the
approach taken in this paper.

It is frequently the case (as in this paper) that the linear equation
is a wave equation. Then, it is well-known that the {\it squares} of
the normal modes frequencies coincide with the spectrum of that linear
self-adjoint operator which is naturally connected with the equation
and the boundary conditions. Since this spectrum is real, the normal
mode frequencies are purely imaginary or real. Using the so called
{\it functional calculus} associated with the operator, a {\it
 representation in terms of the normal modes} can be given for the
solution of the initial-value problem for the linear equation.

{\it Quasinormal mode solutions} of the linear equation are often
displayed by, in some sense, {\it dissipative} systems. They satisfy
boundary conditions which differ from that for the normal modes, but
usually are viewed in the same context as the normal mode solutions.
From this point of view it is natural to ask whether they are in any
sense complete \cite{pricehusain}. On the other hand {\it quasinormal
 frequencies} have, in general, both real and imaginary parts and
hence their {\it squares cannot belong to a spectrum of any linear
 self-adjoint operator}.

In the special case considered in this paper the system is initially
contained in some finite box in space and is {\it``dissipative'', if
one considers the energy contained in the box as a function of
time}. But the system is {\it conservative} if one considers the
energy distributed in the whole space. It turns out that the
quasinormal frequencies of the ``finite'' system are {\it resonances}
of the operator corresponding to the {\it ``infinite''} system. The
analogous can be seen to be true for many other systems.\footnote
{Such resonances are known to be important in quantum theory and
mathematical methods have been developed to deal with them (Volumes
III and IV of \cite{reedsimon}). However it is also known that the
concept of resonances of an operator is far more delicate than that
of the spectrum. In contrast to the spectrum, resonances depend not
only on the operator, but also on the choice of dense subspace of
the underlying function space. In addition much less is known about
resonances than about spectra, concerning in particular their
behaviour under perturbations of the operator.}

This paper addresses the completeness question of the resonance modes
of the infinite system using the framework of ``spectral theory''.
The system is described by a wave equation in one-dimensional space
(as motivated by astrophysical systems). That the system is initially
contained in a finite box is displayed by the fact that only initial
values with compact support are considered.

\section{Introduction}

The decay in time of the solutions of the Einstein field equations
linearized around the Schwarzschild metric is governed by quasinormal
frequencies (``QNF'') and the corresponding modes (``QNM'')
\cite{vishveshwara}. For perturbing fields of the form
\begin{equation} \label{perturbingfield}
\Phi(t,x,\theta,\varphi):=\frac{1}{r} \phi(t,x) \cdot 
{\bf Y}_{\ell m}(\theta,\varphi) \, ,
\end{equation}
where ${\bf Y}_{\ell m}$ denotes an appropriate tensor spherical
harmonic function; $t, r, \theta, \varphi$ are the usual Schwarzschild
coordinate functions; $x:=r+\ln (r-1)$ is the ``tortoise'' coordinate
function and $\ell$ is a natural number, one gets the following wave
equation for the scalar function $\phi$,\footnote {Here the units are
 chosen such that the Schwarzschild radius is normalized to 1.}
\begin{equation}
\frac{\partial^2 \phi}{\partial t^2} +
\left( - \frac{\partial^2}{\partial x^2}+U\right) \phi = 0 \, ,
\label{wave_equation}
\end{equation}
where
\begin{equation}
U(r):= \left(1-\frac{1}{r}\right) \cdot
 \left(\frac{l(l+1)}{r^2}-\frac{3}{r^3}\right) \, .
\label{Schwarzschildpotential}
\end{equation}

A still open mathematical question \cite{pricehusain}, is whether, and
then in which sense, the solutions of (\ref{wave_equation})
corresponding to $C^\infty$-data with compact support can be
represented as sums of quasinormal mode solutions of
(\ref{wave_equation}). The latter are separated solutions satisfying
so called ``purely outgoing'' boundary conditions (see e.g.
\cite{ferrarimashoon}). Since there are an infinite number of such
modes \cite{motetbachelot} it is in particular important to find out
the type of convergence with respect to which such an expansion may be
valid.

The answer to these questions is obscured by technical problems ---
the QNF are not explicitly known and there is no convenient analytical
representation for the QNM.

\begin{figure}[h]
 \def\epsfsize#1#2{0.8#1} \centerline{\epsfbox{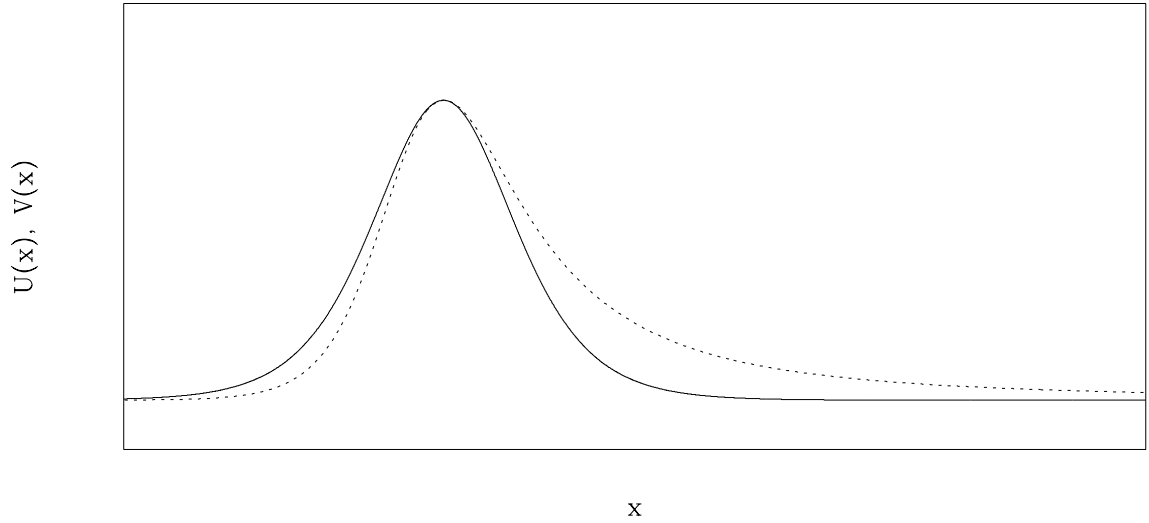}}
\caption{A comparison of the P\"{o}schl-Teller and Schwarzschild potentials.
 The parameters for the P\"{o}schl-Teller potential are fixed by
 setting the maximum amplitude and the second derivative at this
 maximum amplitude to be equal to those for the Schwarzschild
 potential (with $l=2$). That is, $V_0=0.61$ and $b=2.75$. In the
 figure, the solid line shows the P\"{o}schl-Teller, and the dotted
 line the Schwarzschild potential.}
\label{figure1}
\end{figure}

In such a situation it is natural to ask whether there is any reason
to expect that such a quasinormal mode expansion exists? Or more
precisely, is there a wave equation of type (\ref{wave_equation})
having infinitely many quasinormal modes such that each solution
corresponding to $C^\infty$-data with compact support has an expansion
into quasinormal modes? To my knowledge such a wave equation is not
known. Hence it is still unclear whether one should expect such a
``quasinormal mode expansion'' for (\ref{wave_equation}) to exist.
Further, if such a normal mode expansion does exist for
(\ref{wave_equation}) a natural next step would be to ask whether this
is true also for other wave equations, or in other words, whether the
phenomenon is in any sense ``stable'' against `` small perturbations''
of the potential. Such points suggest the consideration of other wave
equations than (\ref{wave_equation}) and in this paper we now look at
the wave equation
\begin{equation} 
\label{wave_equation2}
\frac{\partial^2 \phi}{\partial t^2}+\left(-\frac{\partial^2}
{\partial x^2} +V\right) \phi = 0 \, ,
\end{equation} 
where the potential $V$ is the P\"{o}schl-Teller potential
\cite{poeschlteller},
\begin{equation}
\label{poeschlpotential}
V(x):= \frac{V_{0}} {\cosh^2 (x/b)} , \qquad x \in {\Bbb{R}} \, .
\end {equation} 
Here $V_{0}$ and $b$ are, respectively, the maximal value and the
``width'' of $V$ and are non zero positive real numbers (considered as
given in the following). There are good reasons for working with this
special choice of the so called ``P\"{o}schl-Teller'' potential $V$
instead of the Schwarzschild potential $U$. First, the QNF and QNM are
known analytically \cite{ferrarimashoon}, and there are an infinite
number of QNF which are elementary functions of $V_0$ and $b$. In
addition the shapes of $U$ and $V$ are similar (see Figure
\ref{figure1}) and both potentials are integrable over the real line
and decay exponentially for $x \rightarrow - \infty$. However, the
decay of $U$ and $V$ differs for $x \rightarrow \infty$, where $U$
decays as $1/r^2$ and $V$ decays exponentially. These similarities
have already been used in order to approximate the QNF of the
Schwarzschild black hole which have ``low'' imaginary part by the
corresponding QNF for $V$ \cite{ferrarimashoon}.  A final very
important reason for considering this particular wave equation is that
the resolvent of the Sturm-Liouville operator corresponding to $V$
(given later in Equation~(\ref{operator})) can be given explicitly in
terms of well-known analytic special functions.  This cannot be done,
so far, for the Schwarzschild potential $U$ --- and it is this fact
which prevents the same analysis in this paper being carried through
for (\ref{wave_equation}).

\begin{figure}[h]
 \def\epsfsize#1#2{0.8#1} \centerline{\epsfbox{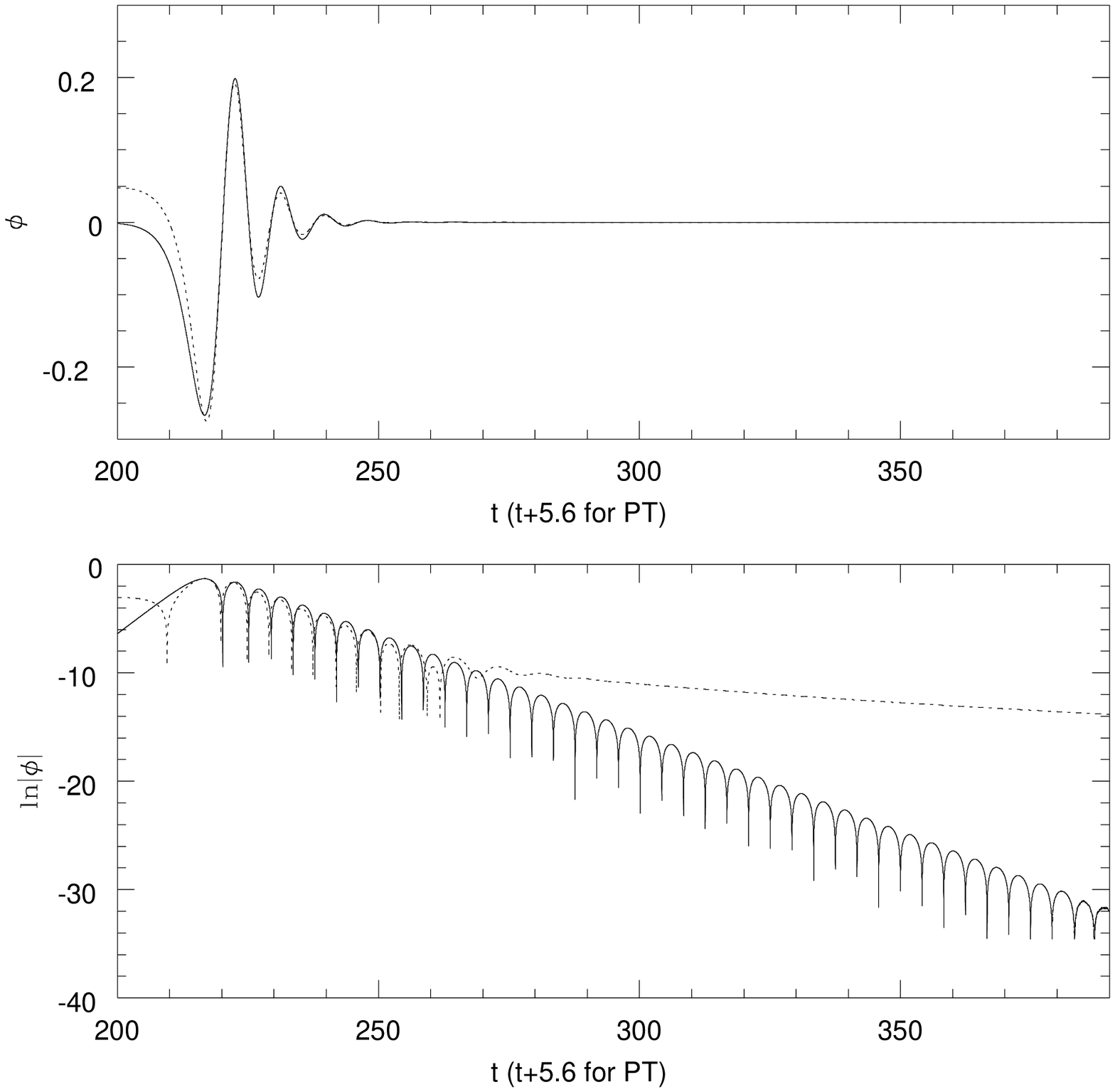}}
\caption{Comparison of the solutions to Equation~(\ref{wave_equation}) 
(with $l=2$) and Equation~(\ref{wave_equation2}) (with $V_0=0.61$ and
 $b=2.75$) from the same initial data ($\phi(0,x) =
 \exp(-1.5(x-120)^2)$, $\phi_{,t}(0,x) = \phi_{,x}(0,x)$.)}
\label{figure2}
\end{figure}

From such considerations it appears that the use of the wave equation
with P\"{o}schl-Teller potential is a good starting point for a
mathematical investigation of the completeness of quasinormal
modes. One may hope that, given the different decay as $x \rightarrow
\infty$ the results have some similarities with those for $U$. This is
illustrated in Figure \ref{figure2}, where the solutions of
(\ref{wave_equation}) and (\ref{wave_equation2}) are compared. In both
cases, the initial data describes a gaussian pulse which is purely
incoming from infinity. In the figure, the lines show the resulting
outgoing waves, as seen by a distant observer. The solid line
corresponds to the P\"{o}schl-Teller potential and the dotted line
corresponds to the Schwarzschild potential. At early times, the
solutions are very similar, although their behaviours differ at late
times.

The most difficult and time-consuming part of the calculations for the
results on completeness, was in the derivation of the estimates,
(\ref{estimate_2}) and (\ref{estimate_3}), on the analytic
continuation of the resolvent of the Sturm-Liouville operator with
P\"oschl-Teller potential. It was not clear {\it a priori}, from
previous works on quasinormal modes, what form the estimates should
take in order to prove or disprove these completeness
results. Although the estimates are given here only for the
P\"oschl-Teller potential, one can hope that their structure is
representative for other potentials. If this is the case, the form of
the estimates (\ref{estimate_2}) and (\ref{estimate_3}) could provide
a basis for further completeness calculations for different
potentials.

Section~3, which contains the rigourous basis of this paper, is intended
to be partly pedagogical. The results apply to a much more general class
than just partial differential operators. Although these results can be found
in the mathematical literature, they are not easily accessible, and in 
this section the relevant results are collected and presented in a
manner more convenient for quasinormal mode considerations.

A study of the literature on quasinormal modes shows that some of these
results (especially (\ref{representation_of_phi_g}) and 
(\ref{representation_2_of_phi_g})) are already used. However, the form 
used is often not valid for the case considered, or the proof of its 
validness is left open. Formulae (\ref{representation_of_phi_g}) and 
(\ref{representation_2_of_phi_g}) in Section~3 offer a rigorous starting 
point for such considerations in the future. Further, in some more 
physically motivated papers dealing with quasinormal mode expansions, 
mathematical terminology such as``convergence" or``completeness", is 
used somewhat freely. That is, the terminology is used but corresponding 
proofs are not given rigorously, or are substituted by ``physical" 
arguements. While important, physical intuition into whether or not an 
infinite sum converges is very different from, and cannot substitute, 
a proof of convergence. Hence, in this paper much importance is placed 
on mathematical rigor.

Now, for those readers who are not concerned with the details of the
various proofs, the main results of the paper are summarised. For
this, denote by $q(A)$ the set of quasinormal frequencies of $V$ and
for each $\omega \in q(A)$ denote by $u_\omega$ the corresponding
quasinormal eigenfunction. In addition let $f$ be some complex-valued
$C^\infty$-function with compact support and let $\phi_f$ be the
corresponding solution of (\ref{wave_equation2}) with initial values
\begin{equation}
 \phi_f(0,x) = 0 \qquad \text{and} \qquad \frac{\partial \phi_f}{\partial t}
(0,x) = f(x) \, ,
 \label{initial_conditions1} 
\end{equation}
for all real $x$. Finally denote by $\phi_{ g,f}$ the following
averaged function obtained from $\phi_f$,
\begin{equation} 
\label{definition_of _phi_g1}
\phi_{g,f}(t):=
\begin{cases}
\int^{+\infty}_{-\infty} g^*(x) \cdot
\phi_f(t,x) dx & \text{for $t \geqslant 0$} \\
0 & \text{for $t < 0$}
\end{cases}
 \, ,
\end{equation}
where $g$ is some complex-valued $C^\infty$-function with compact
support. The main results of this paper are,

\begin{enumerate}
\item{} The quasinormal modes of $V$ are complete, in the sense that
 there is a family of complex numbers $c_\omega$, $\omega \in q(A)$
 (given explicitly in Section~5, see (\ref{coefficients})) such that
 for {\it for a large enough} $t_0$ and for every $t \in [t_0,\infty)$
\begin{equation} \label{result}
 \left(c_\omega \cdot \int\limits_{-\infty}^{+\infty} u_\omega
 (y')f(y')dy' \cdot \int\limits_{-\infty}^{+\infty} g^*(x')u_\omega
 (x')dx' \cdot e^{i \omega t} \right)_{\omega \in q(A)}
\end{equation}
{\it is absolutely summable with sum} $\phi_{g,f}(t)$. {\it So the
 summation of this sequence (using any order of summation) gives the
 quasinormal mode expansion of } $\phi_{g,f}$ {\it for large
 times.}\footnote {Here a remark concerning the role of the test
 function $g$ might be in order. This test function is mainly for
 mathematical convenience. Below there is also given a corresponding
 result on the sum of the sequence, which one gets from (\ref{result})
 by formally substituting $f$ by $\delta (x'-x)$and $g$ by $\delta
 (y'-y)$, respectively, for some $x \in {\Bbb{R}}$ and $y \in
 {\Bbb{R}}$. The corresponding sum is then a Green's function (more
 precisely the so called ``commutator -distribution'') which is
 associated to (\ref{wave_equation2}).} In addition, estimates for the
 possible size of such $t_0$ are given depending on the supports of
 $f$ and $g$. It is shown that $t_0$ can be chosen to be any real
 number which is greater than some explicitly given real number
 $M(g,f)$ (see (\ref{definition_of_d(g,f),m(g,f)_and_M(g,f)}) in
 Section~4).
 
\item{} $\phi_{g,f}$ {\it has an analytic extension}
 $\bar{\phi}_{g,f}$ to the strip $(M(g,f),\infty) \times {\Bbb{R}}$
 and the {\it sequence} (\ref{result}) is {\it uniformly absolutely
 summable} on $[t_0,\infty) \times K_0$ {\it with sum}
 $\bar{\phi}_{g,f}$ for each $t_0 \in (M(g,f),\infty)$ and each
 compact subset $K_0$ of ${\Bbb{R}}$. As a consequence the sequence
 (\ref{result}) can be termwise differentiated to any order on that
 strip and the resulting sequence of derivatives is uniformly summable
 on $[t_0,\infty)\times K_0$ with a sum equal to the corresponding
 derivative of $\bar{\phi}_{g,f}$.
 
\item{} A result shown in Appendix~\ref{appendix2} indicates that the
 QNM sum exists only for large enough times. There it is shown that
 the sequence
 \begin{equation} \label{nonabssum}
\left( c_{\omega} \left[ u_{\omega}(0) \right]^2 \right)_{\omega \in q(A)}
\, , 
\end{equation}
which one gets formally\footnotemark[3] from (\ref{result}) by the
substitutions $t=0$, $f$ by $\delta (x)$ and $g$ by $\delta (y)$, {\it
is not absolutely summable}. {\it Hence for that case there cannot be
associated a sum with (\ref{nonabssum}) which is independent of the
order of the summation.}

\end{enumerate}

The plan for the remaining part of this paper is the following: In
{\it Section}~3 the wave equation (\ref{wave_equation2}) is associated
with the linear self-adjoint (Sturm-Liouville) operator $A$
(\ref{operator}). A representation of the solution of the initial
value problem is given. This representation is found by applying the
members of a special family (parameterized by time) of functions of
$A$ (which are bounded linear operators) on the data (see
e.g. \cite{beyer94}). Using a result of semigroup theory
\cite{hillephillips} these functions are represented by integrals over
the resolvent of $A$ (see (\ref{representation_of_phi_g}) or
(\ref{representation_of_phietat}) in
Appendix~\ref{appendix2}). Because of the analyticity properties of
the resolvent the method of contour integration can be used in {\it
Section}~5. Using the residue theorem the quasinormal frequencies (and
modes) come in since they are (in quantum terminology) common poles of
the analytic continuations of a set of transition amplitudes of the
resolvent (see e.g. \cite{reedsimon}, Volume IV, page 55).  By
explicit estimates on these analytic continuations which are supplied
in {\it Section} 4 it is then shown that the resonance modes are
complete for a large enough time $t_0$. In addition estimates for
$t_0$ are given. These bounds depend on the support of the data.  {\it
Section}~6 gives a discussion of the results.  {\it Appendix}~A
supplies mathematical details to the results of Sections~3, 4 and
5. Finally, for readers better acquainted with the ``Laplace method''
\cite{schmidt} than operator theory, {\it Appendix}~B gives a (not
completely rigorous) derivation for the basic representation
(\ref{representation_of_phi_g}) used in this paper for the solution of
the initial value problem for (\ref{wave_equation2}).

\section{An Initial Value Formalism for the Wave Equation}

\label{formalism}
In order to give (\ref{wave_equation2}) a well-defined meaning one
has, of course, to specify the differentiability properties of $\phi$.
In the following a standard abstract approach for giving such a
specification is used.\footnote {See for example, page 295 in
Volume~II of \cite{reedsimon}. Of course there also other approaches
for such a specification. Usually, all approaches turn out to be
``equivalent'' in that the unique solution of the initial value
problem in one approach can be reinterpreted in such a way that it
coincides with the corresponding one in another approach. The approach
chosen in this paper has the advantage that it leads in a natural way
to eigenfunction expansions and/or quasinormal eigenfunction
expansions of the solution.} The purpose of this section is the
derivation of the representations (\ref{representation_of_phi_g}) and
(\ref{representation_2_of_phi_g}) of the solutions of the
initial-value problem of (\ref{wave_equation2}). {\it These
representations are basic for this paper.} The methods for this
derivation come from semigroup theory and spectral theory.  For the
reader not familiar with these methods, this is rederived in
Appendix~B using the so called ``Laplace method''
(e.g. \cite{schmidt}).
 
Define the Sturm-Liouville operator $A:W^2({\Bbb{R}}) \rightarrow
L^2({\Bbb{R}})$ by
\begin{equation}
 A f:= - f^{\prime\prime} + V f \, ,
\label{operator}
\end{equation}
for each $f \in W^2({\Bbb{R}})$. Here $ L^2({\Bbb{R}})$ denotes the
Hilbert space of complex-valued square integrable functions on the
real line with scalar product $<|>$ defined by
\begin{equation}
<f|g> := \int^{+\infty}_{-\infty}f^*(x)\cdot g(x)dx\, ,
\end{equation}
for all $f,g \in L^2({\Bbb{R}})$; $W^2({\Bbb{R}})$ denotes the dense
subspace of $ L^2({\Bbb{R}})$, consisting of two times
distributionally differentiable elements, and the distributional
derivative is denoted by a prime.  By the Rellich-Kato
theorem\footnote{Theorem~X.12 in Volume~II of
\cite{reedsimon}.}, it follows that $A$ is a densely defined linear 
and self-adjoint operator in $L^2({\Bbb{R}})$ which results from
perturbing the linear self-adjoint operator $A_0$ defined by
\begin{equation} 
\label{unperturbed_operator}
 A_0 := \left(W^2 ({\Bbb{R}}) \rightarrow L^2 
 ({\Bbb{R}}), f \mapsto -f^{\prime\prime}\right) \, ,
\end{equation} 
by the bounded linear self-adjoint operator with the function $V$ 
\footnote{See for example Proposition~1 in Chapter~VIII.3, 
Volume~I of \cite{reedsimon}.} (the so called maximal multiplication
operator corresponding to $V$). Further, the spectrum of $A$ consists
of all positive real numbers (including zero). The proof of this,
which is not difficult, is not given here. The formulation of
(\ref{wave_equation2}) used in the following is given by
\begin{equation}
\ddot{\phi}(t) = -A\phi(t) \, ,
\label{wave_equation3} 
\end{equation}
for each $t \in {\Bbb{R}}$, where $\phi$ is required to be a $C^2$-map
from ${\Bbb{R}}$ into $L^2({\Bbb{R}})$ with values in
$W^2({\Bbb{R}})$, and a dot denotes time
differentiation.\footnote{Hence (\ref{wave_equation2}) is viewed,
similarly as in the case of the Schroedinger equation (but with a
second order time derivative), as an ordinary differential equation
for a curve in a Hilbert space.}  {\it Using only abstract properties
of} $A$, namely its selfadjointness and its positiveness, it follows
from the proposition on page 295 in Volume~II of \cite{reedsimon} and
Theorem~11.6.1 in \cite{hillephillips} (see also Theorem~1 in
Appendix~B) that for each $f\in W^2({\Bbb{R}})$ there is a unique
$\phi_f \in C^2({\Bbb{R}},L^2({\Bbb{R}}))$ with values in
$W^2({\Bbb{R}})$, satisfying the initial conditions
\begin{equation}
 \phi_f(0) = 0 \qquad \text{and} \qquad \dot{\phi}_f(0) = f \, ,
 \label{initial_conditions}
\end{equation}
and that the solution $\phi_f$ has the following representation.
Define 
\begin{equation}
\phi_{g,f}(t):=
 \begin{cases}
 <g | \phi_f(t) > & \text{for $t \geqslant 0$} \\
 0 & \text{for $t < 0$ }
 \end{cases} \, .
 \label{definition_of _phi_g}
\end{equation}
The representation of $\phi_f$ is given by 
\begin{equation} 
\label{representation_of_phi_g}
\phi_{g,f}(t) = \frac{1}{\sqrt{2\pi}} e^{\epsilon t}
\left( F^{-1}{\cal R}_{g,f}( \cdot -i\epsilon)\right)(t)\, ,
\end{equation}
for (Lebesgue-) almost all $t \in {\Bbb{R}}$, where $\epsilon$ is an,
otherwise arbitrary, strictly positive real number; $g$ is an,
otherwise arbitrary, element of $L^2({\Bbb{R}})$; $F$ is the unitary
linear Fourier transformation on\footnote{For the definition see
Chapter~IX in Volume~II of \cite{reedsimon}.} $L^2({\Bbb{R}})$ and
${\cal R}_{g,f}:{\Bbb{R}}\times(-\infty,0)\rightarrow{\Bbb{C}}$ is
defined by
\begin{equation} \label{Rgf}
 {\cal R}_{g,f}(\omega) := <g | R(\omega^2)f> \, ,
\end{equation}
for each $\omega\in{\Bbb{R}}\times(-\infty,0)$. Here
$R: {\Bbb{C}} \setminus [0,\infty) \rightarrow L(L^2({\Bbb{R}}), 
L^2({\Bbb{R}}))$ is the so called {\it resolvent} of $A$, which associates 
to each $\lambda \in {\Bbb{C}} \setminus [0,\infty)$ the inverse of the 
operator $A-\lambda $. $L(L^2({\Bbb{R}}), L^2({\Bbb{R}}))$ denotes the 
linear space of bounded linear operators on $L^2({\Bbb{R}})$.

Note that ${\cal R}_{g,f}(\cdot-i\epsilon)$ is square integrable, as can 
easily be concluded from the bound
\begin{equation}
|{\cal R}_{g,f}(\omega)|\leqslant \frac{\| f \|_2 \cdot \| g \|_2 } 
{\max\{2 |\omega_2 | \cdot |\omega_1|,\omega_2^2-\omega_1^2\}} \, ,
\label{est1}
\end{equation}
which is valid for each $\omega = \omega_1 + i\omega_2 \in {\Bbb{R}}
\times (-\infty,0)$.  {\it This bound requires}\footnote{Spectral
Theorem VIII.5(b) in Volume I of \cite{reedsimon}.} {\it also only the
self-adjointness and positivity of $A$}.

Finally, using a well-known property of the Fourier
transformation\footnote{See e.g. the representation of the Fourier
transformation on page~11 in Volume~II of \cite{reedsimon}.}, it
follows from (\ref{representation_of_phi_g}) that there exists a
subset $N$ of ${\Bbb{R}}$ having Lebesgue measure zero such that for
each $t \in [0,\infty) \setminus N$
\begin{equation} \label{representation_2_of_phi_g}
\phi_{g,f} (t) = \frac{1}{2\pi} \lim_{\nu\rightarrow\infty} 
\int_{-\nu}^\nu e^{it \cdot (\omega-i\epsilon)}{\cal R}_{g,f}
(\omega-i\epsilon)d\omega \, .
\end{equation}

The representations (\ref{representation_of_phi_g}) and
(\ref{representation_2_of_phi_g}) have here been given for the special
case of the P\"{o}schl-Teller potential. In fact, as hinted at in the
above text, (\ref{representation_of_phi_g}) is an application of the
abstract Theorem~1 given at the end of Appendix~B, which is far more
general. The representation (\ref{representation_of_phieta}) given in
that theorem is, for instance, also valid for wave equations in
arbitrary space dimensions.

\section{Analytic Properties of the Resolvent}

\label{resolvent}
Formula (\ref{representation_2_of_phi_g}) is the starting point of a
contour integration, which is performed in Section~5, and eventually
leads to the results on the completeness of the quasinormal modes. The
basis for that contour integration is provided by the estimates
(\ref{estimate_2}), (\ref{estimate_3}) of this section below on the
analytic continuation of ${\cal R}_{g,f}$. The purpose of this section
is mainly to explain these estimates. A sketch of the proofs of these
estimates is given in Appendix~A.

Let $f$ and $g$ be arbitrary, considered as given from now on,
complex-valued $C^2$-functions on ${\Bbb{R}}$ with compact supports.
Then it follows from general analytic properties of resolvents that
the function ${\cal R}_{g,f}$ defined in Equation~(\ref{Rgf}) is an
analytic function on the open lower half-plane.

Now {\it using for the first time}\footnote{Apart from its positivity,
which has already been used in concluding that $A$ is a positive
operator.} {\it the special properties of the} P\"{o}schl-Teller
potential it will be concluded that ${\cal R}_{g,f}$ has an analytic
extension into the closed upper half-plane. In order to see this the
auxiliary function $\bar{{\cal R}}_{g,f}$ is now defined,

Define the set $q(A)$ of {\it ``quasinormal frequencies of} $A$'' by
\begin{equation}
q(A):=\bigcup_{k\in{\Bbb{N}}}\left\{\omega_k^{-},\omega_k^{+}\right\} \, ,
\end{equation}
where for each $k \in {\Bbb{N}}$,
\begin{equation} \label{qnfr}
 \omega_k^{-} := i \cdot (\frac{1}{2} -\alpha +k)/b, \quad
 \omega_k^{+} := i \cdot (\frac{1}{2} +\alpha +k)/b \, ,
\end{equation}
and
\begin{equation}
 \alpha := 
 \begin{cases}
 \sqrt{\frac{1}{4}-b^2 V_0} & \text{for $b^2 V_0 \leqslant
 \frac{1}{4}$} \\
 i \sqrt{b^2 V_0 - \frac{1}{4}} & \text{for $b^2 V_0 >
 \frac{1}{4}$}
 \end{cases}.
 \label{definition_of_qA_and_alpha}
\end{equation}

For each $\omega \in {\Bbb{C}} \setminus q(A)$ the corresponding 
$\bar{{\cal R}}_{g,f} (\omega)$ is defined by 
\begin{equation} \label{representation_of_Rgfomega}
 \bar{{\cal R}}_{g,f} (\omega) = \iint 
 \limits_{{\Bbb{R}}^2} g^*(x)K(\omega,x,y)f(y)\,dx\,dy \, ,
\end{equation}
where for each 
$x,y \in {\Bbb{R}}:$
\begin{equation}
K(\omega,x,y)=-\frac{1}{W(\omega)}
\begin{cases}
u_r(\omega,x)u_l(\omega,y) & \text{for $y \leqslant x$} \\
u_l(\omega,x)u_r(\omega,y) & \text{for $y > x$} 
\end{cases} \, ,
\label{definition_of_kernel_K}
\end{equation}
and for each $\omega \in {\Bbb{C}}, x \in {\Bbb{R}}$:
\begin{align}
u_l(\omega,x):=&\,e^{i\omega x}\cdot\bar{F}\left(\frac{1}{2} -
\alpha,\frac{1}{2}+\alpha,1+ib\omega,\frac{1}
{1+e^{-\frac{2x}{b}}}\right) \, ,
\nonumber \\
u_r(\omega,x) := &\, u_l(\omega,-x) \, ,
\end{align}
and
\begin{align}
 W(\omega) := & \, u_l(\omega,x)(u_r(\omega,\cdot))^{\prime}(x) -
 u_r(\omega,x)(u_l(\omega,\cdot))^{\prime}
 \nonumber\\
 = & -\frac{2}{b} \cdot \frac{1}{\Gamma}
 \left(\frac{1}{2}+\alpha+ib\omega\right)
 \frac{1}{\Gamma}\left(\frac{1}{2}-\alpha+ib\omega\right) \, .
 \label{definition_of_ul,ur_and_W}
\end{align}
Here 
 $\bar{F}:{\Bbb{C}}^3 \times U_1(0) \rightarrow {\Bbb{C}}$ 
is the analytic extension of the function
\begin{equation}
 \left({\Bbb{C}}^2 \times ({\Bbb{C}} \setminus - {\Bbb{N}}) 
 \times U_1(0) \rightarrow {\Bbb{C}}, 
 (a,b,c,z) \mapsto F(a,b,c,z)/ \Gamma(c)\right) \, ,
\end{equation}
 where the hypergeometric function 
(Gauss series) $F$ and the Gamma function $\Gamma$ are defined according to 
\cite{abramowitz} and $ 1/ \Gamma$ denotes the extension of 
$\left({\Bbb{C}} \setminus -{\Bbb{N}} \rightarrow {\Bbb{C}}, c 
\mapsto 1/ \Gamma(c)\right) $ to an entire analytic function.
 
Note that for each $\omega \in {\Bbb{C}}$ the corresponding 
functions $u_l(\omega,\cdot)$, $u_r(\omega,\cdot)$ satisfy,
\begin{align}
 (u_l(\omega,\cdot))''(x) - (V(x)-\omega^2) \cdot u_l(\omega,x) &= 0
 \, ,
 \nonumber\\
 (u_r(\omega,\cdot))''(x) - (V(x)-\omega^2) \cdot u_r(\omega,x) &= 0
 \, ,
 \label{homDgl}
\end{align}
for each $x \in {\Bbb{R}}$. In addition for each $\omega \in {\Bbb{R}}
\times (-\infty,0)$ the associated $u_l(\omega,\cdot)$,
$u_r(\omega,\cdot)$ is ${\cal L}^2$ near $-\infty$ and $+\infty$,
respectively. Using this, along with general results on
``Sturm-Liouville'' operators (see e.g. \cite{weidmann}) and
differentiation under the integral sign, it follows that $\bar{{\cal
R}}_{g,f}$ is an analytic function on ${\Bbb{C}} \setminus q(A)$,
which coincides with ${\cal R}_{g,f}$ on the open lower
half-plane. The proof of this is elementary and not given in this
paper.

The QNF of $A$, which coincide with the zeros of the Wronskian determinant 
function $W$, are poles of $\bar{{\cal R}}_{g,f}$. These poles are simple 
for the case $\alpha \neq 0$ and second order for the case $\alpha = 0$. 
In somewhat misleading, but common mathematical terminology, such poles 
are often called ``second sheet poles of the resolvent (of $A$)'' or 
``resonances'' (of $A$) (see for example Volume IV of \cite{reedsimon}). 
This terminology is somewhat misleading, because they not only depend on 
$A$, but also on the choice of a dense subspace of ${\cal L}^2$ 
(see for example Volume IV of \cite{reedsimon}), which is here the 
space of complex-valued $C^{\infty}$- functions on the real line with 
compact support, which is the space from where the data for 
(\ref{wave_equation2}) are taken. 

The QNM corresponding to the QNF of $A$, 
$u_r(\omega, \cdot) u_l(\omega, \cdot)$, $\omega \in q(A)$ satisfy,
\begin{align}
 u_r\left(\omega_k^{\pm},x\right) =& (-1)^k \cdot
 u_l\left(\omega_k^{\pm},x\right)
 \nonumber \\
 u_{\omega_k^{\pm}}(x) := u_l \left(\omega_k^{\pm},x\right)
 =&\frac{1}{\Gamma(\frac{1}{2}\mp \alpha-k)} \cdot \left(2\cosh(x/b)
 \right)^{\frac{1}{2}\pm\alpha+k} \cdot \\ & F\left(-k, -k \mp 2
 \alpha, \frac{1}{2} \mp \alpha-k, 1/(1+e^{-2x/b})\right) \nonumber
 \, ,
\end{align}
for each $k \in {\Bbb{N}}$, $x \in {\Bbb{R}}$. This result is also 
easy to see and its proof is not given in this paper.

In view of the analytic properties of ${\cal R}_{g,f}$ it is natural to 
try to evaluate the right hand side of (\ref{representation_2_of_phi_g}) 
by contour integration. This is done in the next section and that contour 
integration leads to the completeness results of this paper on the QNM 
of the P\"{o}schl-Teller potential. The basis for the contour integration 
is provided by estimates on $\bar{{\cal R}}_{g,f}$ which are now given.

The estimates depend on the parameters $d(g,f)$, $m(g,f)$ and $M(g,f)$, 
which define certain ``distances'' between the supports of $g$ and $f$. 
These distances are defined by,
\begin{align}
d(g,f) := &\min \{ |x-y|: x \in \text{supp($g$) and} \,\, y \in 
  \text{supp($f$)} \} \, , \nonumber\\
m(g,f) := &\max \{ |x-y|: x \in \text{supp($g$) and} \,\, y \in 
  \text{supp($f$)} \} \, ,\\
M(g,f) := &\max \{ D(x,y): x \in \text{supp($g$) and} \,\, y \in
 \text{supp($f$)} \} \geqslant m(g,f) \, ,
\label{definition_of_d(g,f),m(g,f)_and_M(g,f)}
\nonumber
\end{align}
where
\begin{equation}
 D(x,y) := |x-y|+ b \cdot
 \begin{cases}
 \ln (1+2e^{-2x/b})+\ln(1+2e^{2y/b}) & \text{for $y \leqslant x$} \\
 \ln(1+2e^{-2y/b}) + \ln(1+2e^{2x/b}) & \text{for $y > x$}
 \end{cases} \, ,
 \label{definition_of_D(x,y)}
\end{equation}
for each $x\in{\Bbb{R}}$ and $y \in {\Bbb{R}}$. Note that the 
quantities $d(g,f)$ and $m(g,f)$ have an obvious geometrical interpretation.

The following estimates hold for $\bar{{\cal R}}_{g,f}$ and each 
$\omega=\omega_1+i\omega_2\in{\Bbb{C}}\setminus\left(q(A)\cup-q(A)\right)$
\begin{multline}
 |\bar{{\cal R}}_{g,f} (\omega)| \leqslant \\ C_1 (g,f) \cdot
 \frac{e^{2 \pi b|\omega_1|}} {|\cos(2 \pi \alpha) + \cosh(2 \pi b
 \omega)|} \cdot \left(1+4 b^2 \omega_1 ^2\right)^{-1/2} \cdot
 \begin{cases}
 e^{\omega_2 \cdot d(g,f)} & \text{for $ \omega_2 < 0 $} \\
 e^{\omega_2 \cdot M(g,f)} & \text{for $ \omega_2 \geqslant 0
 $}
 \end{cases} \, ,
 \label{estimate_2}
\end{multline}
and if in addition both $\text{supp($f$)}\subset[0,\infty)$ 
and $\text{supp($g$)}\subset(-\infty,0]$ or 
$\text{supp($f$)}\subset(-\infty,0]$ and  
$\text{supp($g$)} \subset [0,\infty):$
\begin{multline}
 |\bar{{\cal R}}_{g,f} (\omega)| \leqslant
 \\
 C_2 (g,f) \cdot \frac{e^{2 \pi b|\omega_1|}}{|\cos(2 \pi \alpha) +
 \cosh(2 \pi b \omega)|} \cdot \left(1+4 b^2 \omega_1
 ^2\right)^{-1/2} \cdot
 \begin{cases}
 e^{\omega_2 \cdot d(g,f)} & \text{for $ \omega_2 < 0 $}
 \\
 e^{\omega_2 \cdot m(g,f)} & \text{for $ \omega_2 \geqslant 0 $}
 \end{cases},
 \label{estimate_3}
\end{multline}
where $C_1(g,f), C_2(g,f) \in [0,\infty)$ are given in 
Appendix~\ref{appendix1}.

The derivation of (\ref{estimate_2}) and (\ref{estimate_3}) is given in 
Appendix~\ref{appendix1}. They were obtained by different methods of 
estimation. Note that depending on the methods used in their derivation 
these estimates are ``singular'' in the open lower half-plane at the 
elements of $-q(A)$, although $\bar{{\cal R}}_{g,f}$ is analytic there. 
This will not be relevant in the following. From (\ref{estimate_2}) and 
(\ref{estimate_3}) follows, in particular, that the {\it restriction of} 
$\bar{{\cal R}}_{g,f}$ {\it to the real axis is square integrable}. 
This is used in the contour integration in the next section.
Note that the {\it corresponding statement is false for} the 
operator $A_0$ (see (\ref{unperturbed_operator}) for the definition)
although it is only a bounded (i.e. in the operator theoretic sense 
``very small'') perturbation of A. For this case the corresponding 
$\bar{{\cal R}}_{g,f}$ is analytic on $ {\Bbb{C}} \setminus \{ 0 \}$ 
and has in general a first order pole at $\omega = 0$.

\section{Consequences}

A first implication of the estimates 
 (\ref{estimate_2}) and 
 (\ref{estimate_3}) 
is, roughly speaking that, 
 for the special case of the P\"{o}schl-Teller potential, the formula 
 (\ref{representation_of_phi_g})
is also true for the case $\epsilon=0$,
making subsequent contour integration easier.

This can be seen as follows. 
The estimates 
 (\ref{estimate_2}) and 
 (\ref{estimate_3})
imply the boundedness of the function which associates the value
$\| {\cal R}_{g,f}(\cdot + i \omega_2) \|_2$ to each 
$\omega_2 \in (-\infty,0)$, where $\| \quad \|_2 $ denotes the norm
which is induced on $L^2({\Bbb{R}})$ by the scalar product $<| >$. 
Hence it follows by a Paley-Wiener theorem\footnote{see for example 
Theorems~1 and 2 in Section~4, Chapter~VI of \cite{yosida}.} that the 
sequence $({\cal R}_{g,f}(\cdot + i \omega_2))_{\omega_2 \in (-\infty,0)}$
converges for $\omega_2 \rightarrow 0$ in $L^2({\Bbb{R}})$ to the 
restriction $\bar{\cal R}_{g,f} |_{\Bbb{R}}$ of $\bar{\cal R}_{g,f}$
to the real axis. 

Hence (\ref{representation_of_phi_g}) and the continuity of the Fourier 
transformation leads to
\begin{equation}
\phi_{g,f} = \frac{1}{\sqrt{2\pi}} \cdot F^{-1}
\bar{\cal R}_{g,f} |_{\Bbb{R}} \, .
\end{equation}
Using a well-known result in the theory of the Fourier 
transformation\footnotemark[10] it follows that there exists a subset
$N$ of ${\Bbb{R}}$ having Lebesgue measure zero such that for each
$t \in [0,\infty) \setminus N$
\begin{equation}
\phi_{g,f}(t) = \frac{1}{2\pi}\cdot \lim_{\nu\rightarrow\infty}
\int^\nu_{-\nu}e^{it\omega}\cdot\bar{\cal R}_{g,f}(\omega) d\omega \, .
\label{int}
\end{equation}
In particular this implies that
 $\phi_{g,f}$ is square integrable --- the corresponding 
statement not being generally true when the operator $A$ is 
replaced by $A_0$.

Equation~(\ref{int}) can now be contour integrated, 
 using the Cauchy integral theorem and Cauchy integral formula, 
 to give an expansion of $\phi_{g,f}$ with respect to the QNM.
{\it In the following, for convenience, the case $\alpha = 0$
 is excluded.} 
Then the QNF of $A$, are simple poles of $\bar{{\cal R}}_{g,f}$. 
But with the help of (\ref{estimate_2}) and 
 (\ref{estimate_3}) the same contour integration can also be 
 carried through for the case $\alpha = 0$ leading to similar results.

The contours are chosen as the boundaries of the rectangles with corners
\begin{align*}
 &(-\nu,0),\,(\nu,0),\,(\nu,n/b),\,(-\nu,n/b) \qquad \text{and}\\
 &(-\nu,0),\,(\nu,0),\,(\nu,-n/b),\,(-\nu,-n/b)
\end{align*}
where 
 $\nu$
is an integer
 and
 $n$
is a natural number. 
Then, following from
 (\ref{estimate_2}) and
 (\ref{estimate_3}),
the integrals along the paths in the upper and lower half plane
vanish for certain $t$ in the limit when first 
 $\nu \rightarrow \infty$ 
and then
 $n \rightarrow \infty$. 
The calculations for this are elementary, but lengthy, and will
not be carried through in this paper, only their results will be given
in the following.

In particular, as demanded by causality, the function $\phi_{g,f}$
vanishes on the interval $[0,d_{g,f}]$, as is seen by closing the 
contour in the lower half plane.

Closing the contour in the upper half plane leads to two statements 
concerning the expansion of $\phi_{g,f}$ in the QNM. First define 
$\mu$ by 
\begin{equation}
 \mu := 
 \begin{cases}
 m(g,f) & \text{if }
 \left\{
 \begin{array}{cccc}
 \text{either}& \text{supp}(f) \subset [0,\infty)
 &\text{and}&
 \text{supp}(g) \subset (-\infty,0]\\
 \text{or} & \text{supp}(f) \subset (-\infty,0]
 &\text{and}& \text{supp}(g) \subset [0,\infty)
 \end{array}\right.\\
 M(g,f) & \text{otherwise}
 \end{cases}\, .
\end{equation} 
Now define for each $n \in {\Bbb{N}}$ and $t \in {\Bbb{C}}$
the entire analytic function $s_{g,f,n}$ by
\begin{multline}
 {s_{g,f,n}}(t) := \sum_{k=0}^n \left(c_{\omega_k^{-}}
 \int\limits_{-\infty}^{+\infty} u_{\omega_k^{-}}(y)f(y)dy
 \int\limits_{-\infty}^{+\infty} g^*(x)u_{\omega_k^{-}}(x)dx e^{i
 \omega_k^{-} t} \right.
 \\
 + \left. c_{\omega_k^{+}} \int\limits_{-\infty}^{+\infty}
 u_{\omega_k^{+}}(y)f(y)dy \int\limits_{-\infty}^{+\infty}
 g^*(x)u_{\omega_k^{+}}(x)dx e^{i \omega_k^{+} t} \right) \, ,
 \label{qnmodesum}
\end{multline}
where for each $k \in {\Bbb{N}}$
\begin{align} 
 c_{\omega_k^{-}} & := \frac{(-1)^k \pi /
 \left(2\sin(2\pi\alpha)\right)} {\Gamma(1+k) \Gamma(1-2\alpha+k)}
\, ,
 \nonumber \\
 c_{\omega_k^{+}} & := \frac{(-1)^{k+1} \pi /
 \left(2\sin(2\pi\alpha)\right)} {\Gamma(1+k) \Gamma(1+2\alpha+k)}
 \, .
 \label{coefficients}
\end{align}
The following statements (i) and (ii) are then true.
\begin{description}
\item{(i)} For each $t_0 \in (\mu,\infty)$ the sequence
 $(s_{g,f,n})_{n\in{\Bbb{N}}}$ converges on $[t_0,\infty)$ in the
 ${\cal L}^2$-mean to $\phi_{g,f}$.
\item{(ii)} The restriction of $\phi_{g,f}$ to $(\mu,\infty)$ has an
 extension to an analytic function on the strip $(\mu,\infty) \times
 {\Bbb{R}}$. For each $t_0 \in (\mu,\infty)$ and each compact subset
 $K_0$ of ${\Bbb{R}}$ the sequence $(s_{g,f,n})_{n\in{\Bbb{N}}}$
 converges uniformly on $[t_0,\infty)\times K_0$ to this extension.
\end{description}

Note that in these results a special order of the summation for the 
QNM sequence (\ref{qnmodesum}) is used, which is induced by the chosen 
contour in the integration. That this result is independent of this 
order of summation for $\mu := M(g,f)$ follows from further results 
on the summability of the sequence
\begin{equation} \label{sequence}
 \left( c_{\omega} u_{\omega}(y) u_{\omega}(x) e^{i \omega t} 
 \right)_{\omega \in q(A)} \, ,
\end{equation}
for given
$x \in {\Bbb{R}}$, $y \in {\Bbb{R}}$ and $t\in [0,\infty)$, which are 
now stated. The corresponding proofs are given in Appendix \ref{appendix2}.
There, it is shown by direct estimates on the sequence elements that, 
given $x\in {\Bbb{R}}$ and $y \in {\Bbb{R}}$, this sequence is 
{\it absolutely and uniformly summable} on $[t_0,\infty) \times K_0$
where $t_0 > D_s(x,y)$, $K_0$ is any compact subset of ${\Bbb{R}}$ and 
\begin{equation} \label{condition}
 D_s(x,y) :=b \log\left(2\left[\cosh\left(\frac{x-y}{b}\right) + 
 \cosh\left(\frac{x+y}{b}\right)\right] \right) \, .
\end{equation} 
Hence, in particular follows the analyticity of the function which 
associates to each 
$t \in (D_s(x,y),\infty) \times {\Bbb{R}}$ the value
\begin{equation}
\sum_{\omega \in q(A)}
c_\omega u_\omega (y)u_\omega (x) e^{i \omega t} \, .
\end{equation}
Further it is shown in Appendix \ref{appendix2} that 
\begin{equation} \label{qnsequences}
 \left(c_\omega \cdot \int\limits_{-\infty}^{+\infty} u_\omega (y)f(y)dy
 \cdot \int\limits_{-\infty}^{+\infty} g^*(x)u_\omega (x)dx 
 \cdot e^{i \omega t} \right)_{\omega \in q(A)} \, ,
\end{equation}
is {\it absolutely and uniformly summable} on $[t_0,\infty) \times
K_0$, where $t_0 > M_s(g,f)$, $K_0$ is any compact subset of
${\Bbb{R}}$ and
\begin{equation} \label{Msgf}
 M_s(g,f) := \max \{ D_s(x,y): x \in \text{supp($g$)} \;\; \text{and} 
 \;\; y \in \text{supp($f$)} \} \, .
\end{equation}

It is easily seen that
\begin{equation} \label{DDs}
 D(x,y) \geqslant D_s(x,y) \, ,
\end{equation} 
for all $x \in {\Bbb{R}}$ and $y \in {\Bbb{R}}$ and hence that
\begin{equation} \label{MMs}
 M(g,f) \geqslant M_s(g,f) \, .
\end{equation} 
Using the results on the QNM sequence from this section above, it
follows that for every $t_0 > M(g,f)$ and for every $t \in
[t_0,\infty)$ the sequence (\ref{qnsequences}) {\it is absolutely
summable with sum } $\phi_{g,f}(t)$ and that the {\it sequence}
(\ref{qnsequences}) is {\it uniformly absolutely summable} on
$[t_0,\infty) \times K_0$ {\it with sum} $\bar{\phi}_{g,f}$ for each
$t_0 \in (M(g,f),\infty)$ and each compact subset $K_0$ of
${\Bbb{R}}$. As a consequence the sequence (\ref{qnsequences}) can be
termwise differentiated to any order on that strip and the resulting
sequence of derivatives is uniformly summable on $[t_0,\infty)\times
K_0$ with a sum equal to the corresponding derivative of
$\bar{\phi}_{g,f}$.

A further result shown in Appendix~\ref{appendix2} indicates that the
QNM sum exists only for large enough times. There it is shown that,
for the special case of $x=y=0$ and $t=0 \left(< D_s(0,0)\right)$ the
sequence (\ref{sequence}) {\it is not absolutely summable} because the
sum
\begin{equation} 
 \label{sequence2}
 \sum_{k=0}^n \left| c_{\omega_k^{-}} 
 \left[u_{\omega_k^{-}}(0)\right]^2 \right| \, ,
\end{equation}
is shown to {\it diverge} for $n \to \infty$.  {\it Hence, for that
case, there can be no associated sum with (\ref{sequence}) which is
independent of the order of the summation.}

\section{Discussion and Open Questions}

\label{discussion}

In this paper we gave several results on the completeness of the
quasinormal modes of the P\"{o}schl-Teller potential. A main result is
that any solution of the wave equation with the P\"{o}schl-Teller
potential (\ref{wave_equation2}) corresponding to $C^{\infty}$-data
with compact support can be expanded uniformly in time with respect to
the quasinormal modes after a large enough time $t_0$. Further the
corresponding series are absolutely convergent, and hence do not
depend on the order of summation. In addition we showed that these
series can be arbitrarily often termwise partially differentiated with
respect to time, again leading to series which converge absolutely and
uniformly in time on $[t_0,\infty)$ to the corresponding time
derivatives of the solution. Estimates of $t_0$ were given which
depend on the support of the data and on the point of observation.

Estimates were also given for the time $t_1$ from when the solution
can be expanded uniformly in time with respect to the quasinormal
modes, {\it where a special order of summation is assumed}. Also for
this case the quasinormal mode series can be arbitrarily often
termwise partially differentiated with respect to time thereby leading
to series which converge uniformly in time on $[t_1,\infty)$ to the
corresponding time derivatives of the solution of the initial value
problem.

These estimates have in common that they depend on both the support
properties of the data and the point of observation, and that they are
greater or equal to the geometrical distance between the support of
the data and the observational point.

We showed that, for an ``early'' time and zero distance between the
support of the data and observational point, the corresponding
quasinormal mode series is {\it not} absolutely convergent. Hence,
there is no associated sum, since in general different orders of
summation will give different results.

Several open questions remain. The results of this paper suggest a
relationship between the convergence of the quasinormal mode sums of
the P\"{o}schl-Teller potential and causality. To make this clearer,
one would like to have a complete overview of the convergence of the
quasinormal mode sums depending on the support of the data as well as
the point of observation; possibly depending on whether a special
order of summation is assumed or not and possibly depending on whether
the series converges to the corresponding solution of the wave
equation or not.

\subsection*{Acknowledgements}

I would like to thank Bernd Schmidt for bringing this problem to my
attention, and for many interesting and helpful discussions about the
subject. I would also like to thank Gabrielle Allen for providing the
figures for the paper, and for her help in the writing of the paper.
Finally, I would like to thank Rachel Capon for useful discussions on
quasinormal modes of the Schwarzschild black hole, and Kostas Kokkotas
for explaining w-modes to me.

\appendix

\section{}
\label{appendix1}

This appendix gives a derivation of the estimates (\ref{estimate_2}) and 
(\ref{estimate_3}) as well as an estimate on the members of the sequence 
(\ref{sequence}). All these estimates are crucial for the proof of the 
expansion formulae in Section~4. The definitions and the notation of 
\cite{abramowitz} are used throughout. The derivation uses the following 
auxiliary estimate.

\noindent
 {\it Lemma 1}. Let $n \in {\Bbb{N}}$, $ a \in (0,1) $ and 
$s \in [0,\infty)$ be given. Then
\begin{equation}
\int_0^{\pi /2} e^{-st}\sin ^{n+a-1} (t)dt \leqslant \frac{\pi B_a}{2}
\cdot \frac{2^{-n}
\Gamma (n+1)}{(\Gamma(\frac{n}{2}+1))^2} \cdot (1+s^2)^{-a/2} \, ,
\label{intest}
\end{equation}
where
\begin{equation}
 B_a := \left(\frac{4}{\pi}\right)^a \cdot \max \left\{2^{-a} \cdot 
 \left(\pi + \frac{1}{a}\right), 
 \Gamma(a)+
 \pi \cdot \left(\frac{a}{e}\right)^a \right\} .
 \label{definition_of_Ba}
\end{equation}
Note that later on, (\ref{intest}) 
has to provide a proper estimate for the vanishing of the integral both 
for $s \rightarrow \infty$ and $n \rightarrow \infty$. This demand excludes, 
for instance, an application of the method of partial integration, in
the following proof.

\noindent
{\it Proof}. First by standard estimates for the sine-function one gets
\begin{align}
 \int_0^{\pi /2}& e^{-st} \sin ^{n+a-1} (t)dt
 \nonumber\\
 &\leqslant \left(\frac{2}{\pi}\right)^{a-1} \cdot \int_0^{\pi /2}
 e^{-st} t^{a-1} \sin ^n (t)dt
 \nonumber\\
 &\leqslant \left(\frac{2}{\pi}\right)^{a-1} \cdot \left[2^{-n} \cdot
 \int_0^{1/2} e^{-st} t^{a-1} dt + 2^{1-a} \cdot e^{-s/2} \cdot
 \int_{1/2}^{\pi /2} \sin ^n (t)dt\right]
 \label{estimates1}\\
 &\leqslant \left(\frac{2}{\pi}\right)^{a-1} \cdot \int_0^\pi \sin ^n
 (t)dt \cdot \left[\frac{1}{\pi} \cdot \int_0^{1/2} e^{-st} t^{a-1}
 dt + 2^{-a} \cdot e^ {-s/2}\right]
 \nonumber\\
 & = \left(\frac{2}{\pi}\right)^{a-1} \cdot \frac{2^{-n}\Gamma
 (n+1)}{(\Gamma(\frac{n}{2}+1))^2} \cdot \left[\int_0^{1/2} e^{-st}
 t^{a-1} dt + \pi \cdot 2^{-a} \cdot e^{-s/2}\right] \nonumber \, ,
\end{align}
where in the last equality the identity
\begin{equation}
 \int_0^\pi \sin ^n (t)dt = 
 \frac{\pi \cdot \Gamma (n+1)}{2^n \cdot (\Gamma(\frac{n}{2}+1))^2}\, ,
 \label{identity1}
\end{equation}
was used\footnote{This can be derived, for instance, using Formulae
 6.2.1, 6.2.2, 6.1.8 and 6.1.18 of 
 \cite{abramowitz}.}.
For the case $0 \leqslant s \leqslant 1$ one has now,
\begin{equation}
\int_0^{1/2} e^{-st} t^{a-1} dt + \pi \cdot 2^{-a} \cdot e^{-s/2}
\leqslant \int_0^{1/2} t^{a-1} dt + \pi \cdot 2^{-a} 
\leqslant \left(\pi + \frac{1}{a}\right) \cdot (1+s^2)^{-a/2} \, ,
\label{estimates2}
\end{equation}
and for the case $s > 1 $,
\begin{align}
 \int_0^{1/2} e^{-st} t^{a-1} dt + \pi \cdot 2^{-a} \cdot e^{-s/2}
 &\leqslant
 s^{-a} \cdot [\Gamma(a) + \pi \cdot 2^{-a} \cdot s^a e^{-s/2}] \nonumber\\
 & \leqslant s^{-a} \cdot \left[\Gamma(a) + \pi \cdot
 \left(\frac{a}{e}\right)^a \right]
\label{estimates3}\\
& \leqslant 2^a \cdot \left[\Gamma(a) + \pi \cdot
 \left(\frac{a}{e}\right)^a \right] \cdot (1+s^2)^{-a/2} . \nonumber
\end{align}
The result (\ref{intest}) then follows from (\ref{estimates1}), 
(\ref{estimates2}) and (\ref{estimates3}).
$_\blacksquare$ 

The starting point for the derivation of the formulae 
(\ref{estimate_2}) and (\ref{estimate_3}) is the following.

\noindent
 {\it Lemma 2} . Let $\omega \in {\Bbb{C}} \backslash (q(A) \cup -q(A))$, 
$x \in {\Bbb{R}}$ and $y \in {\Bbb{R}}$ be given then
\begin{multline}
 K(\omega,x,y) = \frac{\pi^2 b e^{-i\omega \cdot |x-y|}} {\cos(2 \pi
 \alpha)+ \cosh(2 \pi b \omega)} \cdot
 \\
\begin{cases} 
 h(\omega,\alpha,(1+e^{2x/b})^{-1}) \cdot
 h(\omega,-\alpha,(1+e^{-2y/b})^{-1}) &
 \text{for $y \leqslant x $} \\
 h(\omega,-\alpha,(1+e^{-2x/b})^{-1}) \cdot
 h(\omega,\alpha,(1+e^{2y/b})^{-1}) & \text{for $y > x $}
 \end{cases} \, ,
 \label{representation_of_K}
\end{multline}
where for arbitrary $\beta \in (-1/2,1/2) \times {\Bbb{R}}$ and 
$x' \in {\Bbb{R}}$,
\begin{equation}
h(\omega,\beta,x') := 
 \frac{\bar{F}(\frac{1}{2} - \beta , \frac{1}{2} + \beta, 1+ib \omega, x')}
 {\Gamma(\frac{1}{2} + \beta - ib \omega)} \, ,
\label{definition_of_h}
\end{equation}

\noindent
{\it Proof}. The proof consists of a straightforward calculation starting
from (\ref{definition_of_kernel_K}) and using 
(\ref{definition_of_ul,ur_and_W}) in addition to Equations 6.1.17 and 
15.1.1 of \cite{abramowitz}. $_\blacksquare$

Note that the main reason for representing $K(\cdot,x,y)$ in the form 
(\ref{representation_of_K}) is that only the first elementary factor 
is singular at the elements of $q(A)$. The price for this is that this 
factor is singular also at the points of $-q(A)$ in the open upper 
half-plane. But this will play no role in the following.

The function $h$ satisfies the following estimate, which eventually leads 
to (\ref{estimates3}).

\noindent
{\it Lemma 3}. Let $\beta \in {\Bbb{C}}$ with $ -1/2 < Re(\beta) <1/2$, 
$\omega = \omega_1 + i \omega_2 \in {\Bbb{C}}$ such that $\omega \neq in/b$ 
for all $n \in {\Bbb{N}}\setminus \{0\}$ as well as 
$\omega \neq -i \cdot (n+\frac{1}{2}+\beta)/b $ for all 
$n \in {\Bbb{N}}\setminus \{0\}$ and $x \in (0,\frac{1}{2}) $ be given. 
Then,
\begin{equation}
 |h(\omega,\beta,x)| \leqslant C_\beta \cdot (1+4b^2 \omega_1 ^2)^{-\frac{1}{2} \cdot
 (\frac{1}{2} + Re(\beta))} \cdot e^{\pi b |\omega_1|} \cdot (1-2x)^{-(\frac{1}{2} - 
 Re(\beta))} ,
 \label{estimate2h}
\end{equation}
where
\begin{equation}
 C_\beta := \pi ^{-1} \cdot 2^{-\frac{1}{2}+Re(\beta)} \cdot \Gamma\left(\frac{1}{2}
 -Re(\beta)\right) \cdot |cos(\pi \beta)| \cdot e^{\pi \cdot |Im(\beta)|/2} \cdot B_{\frac{1}{2}
 +Re(\beta)} .
 \label{definition_of_Cbeta}
\end{equation}

\noindent
{\it Proof}. First, using the power series expansion of the hypergeometric 
function and Equation~6.1.22 of \cite{abramowitz}, one gets
\begin{align}
 & |h(\omega,\beta,x)| = \left|\frac{1}{\Gamma(\frac{1}{2}+\beta -
 ib\omega) \cdot \Gamma(1+ib\omega)} \cdot \sum_{n=0}^\infty
 \frac{\left(\frac{1}{2}-\beta \right)_n \cdot
 \left(\frac{1}{2}+\beta \right)_n}{(1+ib\omega)_n} \cdot
 \frac{x^n}{\Gamma(n+1)}\right|
 \nonumber \\
 &\leqslant \frac{1}{|\Gamma(\frac{1}{2}+\beta)|} \cdot
 \sum_{n=0}^\infty
 \left|\frac{\Gamma(n+\frac{1}{2}+\beta)}{\Gamma(n+1+ib\omega) \cdot
 \Gamma(\frac{1}{2}+\beta-ib\omega)}\right| \cdot
 \left|\left(\frac{1}{2} -\beta \right)_n \cdot
 \frac{x^n}{\Gamma(n+1)}\right| .
 \label{estimate2h1}
\end{align}
Using the Formula\footnote{See e.g. Equation~5.25 in \cite{oberhettinger}.}
\begin{align}
 \int_{-\pi/2}^{\pi/2} e^{iyt} \cdot \cos^{u-1}(t)dt &=
 e^{i\pi y/2} \cdot \int_{0}^{\pi} e^{-iyt} \cdot \sin^{u-1}(t)dt \nonumber\\
 &= \frac{\pi \cdot 2^{1-u} \cdot \Gamma(u)} {\Gamma(\frac{1+u-y}{2})
 \cdot \Gamma(\frac{1+u+y}{2})},
 \label{aux1}
\end{align}
which is valid for arbitrary $u \in (0,\infty) \times {\Bbb{R}}$ and 
$y \in {\Bbb{C}}$ (where the expression which includes the Gamma functions 
is defined by analytic continuation for the cases $y= \pm (1+u)$ ), one 
gets in a second step,
\begin{align}
 \phantom{1}&
 \left|\frac{\Gamma(n+\frac{1}{2}+\beta)}{\Gamma(n+1+ib\omega) \cdot
 \Gamma(\frac{1}{2}+\beta-ib\omega)}\right| \nonumber \\
 & \leqslant \frac{1}{\pi} \cdot 2^{n-\frac{1}{2} + Re(\beta)} \cdot
 \int_{-\pi/2}^{\pi/2} e^{(2b\omega_1 - Im(\beta)) \cdot t} \cdot
 \cos^{n-\frac{1}{2}+Re(\beta)}(t)dt \nonumber\\
 &\leqslant \frac{1}{\pi} \cdot 2^{n+\frac{1}{2} + Re(\beta)} \cdot
 e^{\pi \cdot |Im(\beta)|/2} \cdot \int_{-\pi/2}^{0} e^{-2b|\omega_1|
 \cdot t} \cdot \cos^{n-\frac{1}{2}+Re(\beta)}(t)dt
 \label{aux2}\\
 &=\frac{1}{\pi} \cdot 2^{n+\frac{1}{2} + Re(\beta)} \cdot e^{\pi
 \cdot |Im(\beta)|/2} \cdot e^{\pi b |\omega_1|} \cdot
 \int_{0}^{\pi /2} e^{-2b|\omega_1| \cdot t} \cdot
 \sin^{n-\frac{1}{2}+Re(\beta)}(t)dt
 \nonumber\\
 &\leqslant 2^{n-\frac{1}{2} + Re(\beta)} \cdot e^{\pi \cdot
 |Im(\beta)|/2} \cdot B_{\frac{1}{2}+Re(\beta)} \cdot (1+4b^2
 \omega_1 ^2)^{-\frac{1}{2} \cdot \left(\frac{1}{2} +
 Re(\beta)\right)} \cdot e^{\pi b |\omega_1|} \, . \nonumber
\end{align}
With help from Equations 6.1.22 and 6.1.26 of \cite{abramowitz} 
one has for an arbitrary $n \in {\Bbb{N}}$,
\begin{align}
 \left|\left(\frac{1}{2} -\beta \right)_n\right| =
 \left|\frac{\Gamma(n+\frac{1}{2}-\beta)}{\Gamma(\frac{1}{2}-\beta)}\right|
 &\leqslant
 \frac{\Gamma(n+\frac{1}{2}-Re(\beta))}{|\Gamma(\frac{1}{2}-\beta)|}
 \nonumber\\ & =
 \frac{\Gamma(\frac{1}{2}-Re(\beta))}{|\Gamma(\frac{1}{2}-\beta)|}
 \cdot \left(\frac{1}{2}-Re(\beta)\right)_n \, .
\label{aux3}
\end{align}
Finally, (\ref{estimate2h}) follows from (\ref{estimate2h1}), 
(\ref{aux2}), (\ref{aux3}) with the help
 of Formulae 6.1.17 and 3.6.8 of \cite{abramowitz}. $_\blacksquare$

From (\ref{representation_of_K}), (\ref{estimate2h}) and the continuity 
of $K$ one gets now the following estimate. 

\noindent
{\it Lemma 4}. Let $\omega \in {\Bbb{C}} \setminus \left(q(A) \cup -q(A) \right)$,
$x \in {\Bbb{R}}$ and $y \in {\Bbb{R}}$ be given. Then,
\begin{align}
 |K(\omega,x,y)| \leqslant & \pi^2 bC_\alpha C_{-\alpha} \cdot
 \frac{e^{2\pi b |\omega_1|}} {|\cos(2 \pi \alpha) + \cosh(2 \pi b
 \omega)|} \cdot (1+4b^2 \omega_1 ^2)^{-\frac{1}{2}} \cdot
 e^{\omega_2 \cdot |x-y|} \cdot \nonumber\\
 &
 \begin{cases}
 (\tanh(x/b))^{-\left(\frac{1}{2} - Re(\alpha)\right)} \cdot
 (\tanh(-y/b))^{-\left(\frac{1}{2} + Re(\alpha)\right)}
 & \text{if $x>0$ and $y<0$} \\
 (\tanh(-x/b))^{-\left(\frac{1}{2} + Re(\alpha)\right)} \cdot
 (\tanh(y/b))^{-\left(\frac{1}{2} - Re(\alpha)\right)} &
 \text{if $x<0$ and $y>0$}
 \end{cases} \, . 
\label{estimateK2}
\end{align}
From this one gets easily (\ref{estimate_3}) (compare (\ref{estimate_3}) and in particular
the assumptions on $f$ and $g$), where
\begin{equation}
 C_2(g,f) := \pi^2 bC_\alpha C_{-\alpha} \cdot
 \iint \limits_{{\Bbb{R}}^2} g^* (x)H_2(x,y)f(y)dxdy \, ,
\end{equation}
and where
\begin{equation}
H_2(x,y) := \begin{cases}
 (\tanh(x/b))^{-\left(\frac{1}{2} - Re(\alpha)\right)} \cdot 
 (\tanh(-y/b))^{-\left(\frac{1}{2} + Re(\alpha)\right)} 
 & 
 \text{if $x>0$ and $y<0$} 
 \\
 (\tanh(-x/b))^{-\left(\frac{1}{2} + Re(\alpha)\right)} \cdot 
 (\tanh(y/b))^{-\left(\frac{1}{2} - Re(\alpha)\right)} 
 & 
 \text{if $x<0$ and $y>0$} 
 \\
 0 
 & 
 \text{otherwise}
 \end{cases}.
\label{definition_of _H2}
\end{equation}

A further estimate of the function $h$ uses an integral representation of 
the hypergeometric function $F$, which could not be found in the tables 
on special functions. For this reason that representation and its proof 
is given now.

\noindent
{\it Lemma 5}. Let $a \in {\Bbb{C}}$, $b \in {\Bbb{C}}$, 
$c \in {\Bbb{C}} \setminus (-{\Bbb{N}})$ and $z \in {\Bbb{C}}$ with 
$|z|<1$ be given. Then
\begin{description}
\item{(i)} if in addition $Re(c) > Re(b)$ and $b \not\in {\Bbb{N}} 
\setminus \{0\}$ hold,
\begin{multline}
 F(a,b,c,z) = \pi ^{-1} \cdot 2^{c-b-1} \cdot e^{i \pi (c+b-1)/2}
 \cdot
 \frac{\Gamma(c) \cdot \Gamma(1-b)}{\Gamma(c-b)} \cdot \\
 \int_{0}^{\pi} e^{-i \cdot (c+b-1)\cdot t} \cdot \sin^{c-b-1}(t)
 \cdot \left(1-ze^{-2it}\right)^{-a}dt \, .
 \label{hyper1}
\end{multline}
\item{(ii)} if in addition $Re(b)>0$ and $c-b \not\in {\Bbb{N}} 
\setminus \{0\}$ hold,
\begin{multline}
 F(a,b,c,z) = \pi ^{-1} \cdot 2^{b-1} \cdot e^{i \pi (2c-b-1)/2}
 \cdot \frac{\Gamma(c) \cdot \Gamma(b-c+1)}{\Gamma(b)} \cdot
 (1-z)^{c-(a+b)} \cdot \\
 \int_{0}^{\pi} e^{-i \cdot (2c-b-1)\cdot t} \cdot \sin^{b-1}(t)
 \cdot \left(1-ze^{-2it}\right)^{a-c}dt \, .
 \label{hyper2}
\end{multline}
\end{description}

\noindent
{\it Proof}. Part~(ii) is a direct consequence of (i) and Formula~15.3.3 
in \cite{abramowitz}. Hence it remains to prove part (i). For this let 
$Re(c) > Re(b)$ and $b \not\in {\Bbb{N}} \setminus \{0\}$. First by 
Formulae~6.1.22 and 6.1.17 in \cite{abramowitz} 
as well as by some elementary reasoning it follows that
\begin{equation}
 \frac{(b)_n}{(c)_n} = (-1)^n \cdot \frac{\Gamma(c) \cdot 
 \Gamma(1-b)}{\Gamma(c-b)} \cdot
 \frac{\Gamma(c-b)}{\Gamma(c+n) \cdot \Gamma(1-(b+n))} \, ,
 \label{auxi1}
\end{equation}
where the right hand side is defined by analytic continuation 
(and hence by zero) for the cases where $b \in \{-n+1,-n+2, ...\}$.
From the definition of $F$ (Formula~15.1.1 in \cite{abramowitz}), and 
using (\ref{aux1}) and (\ref{auxi1}) follows,
\begin{align}
 F(a,b,c,z) =& \frac{\Gamma(c) \cdot \Gamma(1-b)}{\Gamma(c-b)} \cdot
 \sum_{n=0}^\infty \frac{\Gamma(c-b)}{\Gamma(c+n) \cdot
 \Gamma(1-(b+n))} \cdot
 (a)_n \cdot \frac{(-z)^n}{\Gamma(n+1)} \nonumber\\
 = & \pi ^{-1} \cdot 2^{c-b-1} \cdot e^{i \pi (c+b-1)/2} \cdot
 \frac{\Gamma(c) \cdot \Gamma(1-b)}{\Gamma(c-b)} \cdot
 \label{auxi2}\\
 &\lim_{N \rightarrow \infty} \int_{0}^{\pi}e^{-i\cdot (c+b-1) \cdot
 t} \cdot \sin^{c-b-1}(t) \cdot \left(\sum_{n=0}^N
 \frac{(a)_n}{\Gamma(n+1)} \cdot (e^{-2it} \cdot z)^n \right)dt \, .
 \nonumber
\end{align}
From this (\ref{hyper1}) follows using Lebesgue's dominated 
convergence theorem and 
the complex version of Formula 3.6.9 (``binomial series'') of 
\cite{abramowitz}. $_\blacksquare$

Note that part (i) of the foregoing Lemma 5 gives an integral 
representation for the hypergeometric series for a larger class 
of parameter values than Formula 15.3.1 in \cite{abramowitz}, 
since it does not assume that $Re(b)>0$ holds. This will be essential 
for the derivation of (\ref{estimate_2}).

Actually used in the following is the subsequent corollary of Lemma 5,
 
\noindent
{\it Corollary 6}. Let $a \in {\Bbb{C}}$, $b \in (0,\infty) \times 
{\Bbb{R}}$, $c \in {\Bbb{C}} \setminus (-{\Bbb{N}})$ such that 
$c-b \not\in {\Bbb{N}} \setminus \{0\}$ and $x \in (-1,1)$ be given. Then
\begin{align}
 F(a,b,c,x) & = \pi ^{-1} \cdot 2^{b-1} \cdot \frac{\Gamma(c) \cdot
 \Gamma(b-c+1)}{\Gamma(b)} \cdot
 (1-x)^{c-(a+b)} \cdot \nonumber\\
 &\int_{-\pi/2}^{\pi/2} e^{-i \cdot (2c-b-1)\cdot t} \cdot
 \cos^{b-1}(t) \cdot \left(x+e^{2it}\right)^{a-c}dt \, .
\label{hyper3}
\end{align}

\noindent
 {\it Proof}. The relation (\ref{hyper3}) follows from (\ref{hyper2}) 
by a straightforward substitution and from the identity 
\begin{equation}
(1+xe^{-2it})^{a-c} = e^{-2i \cdot (a-c) \cdot t} \cdot 
\left(x+e^{2it}\right)^{a-c} \, ,
\label{auxc}
\end{equation}
for each $t \in (-\pi/2,\pi/2)$. The latter can easily be shown by 
analytic continuation. $_\blacksquare$

Now with the help of these auxiliary results a further estimate for the function $h$ will be 
proved, which eventually leads to (\ref{estimate_2}). 

\noindent
 {\it Lemma 7}. Let $\beta \in {\Bbb{C}}$ with 
$ -1/2 < Re(\beta) <1/2$, $\omega = \omega_1 + i \omega_2 \in {\Bbb{C}}$ such that 
$\omega \neq in/b $ for all $n \in {\Bbb{N}}\setminus \{0\}$ as well as $ \omega \neq -i \cdot 
(n+\frac{1}{2}+\beta)/b $ for all $n \in {\Bbb{N}}\setminus \{0\}$ and $x \in (0,1) $ be
given. Then,
\begin{align}
 |h(\omega,\beta,x)| \leqslant &C'_\beta \cdot (1+4b^2 \omega_1
 ^2)^{-\frac{1}{2} \cdot (\frac{1}{2} + Re(\beta))} \cdot e^{\pi b
 |\omega_1|} \cdot (1-x)^{-(\frac{1}{2} +
 Re(\beta))} \cdot \nonumber\\
 &\begin{cases}
 \left(\frac{1+x}{1-x}\right)^{b\omega_2} & \text{for $\omega_2 > 0 $} \\
 1 & \text{for $\omega_2 \leqslant 0$} 
 \end{cases}\, ,
\label{estimate1h}
\end{align}
where
\begin{equation} 
C'_\beta := \frac{2^{-\frac{1}{2}+Re(\beta)} \cdot e^{5\pi \cdot |Im(\beta)|/2}}
 {|\Gamma(\frac{1}{2}+\beta)|} \cdot B_{\frac{1}{2}+Re(\beta)} .
\label{definition_of_Cpbeta}
\end{equation}
{\it Proof}. First one gets from the definitions and (\ref{hyper3}),
\begin{align}
 |h(\omega,\beta,x)| = &\frac{\pi^{-1} \cdot
 2^{Re(\beta)-\frac{1}{2}}} {|\Gamma(\frac{1}{2}+\beta)|} \cdot
 (1-x)^{-b\omega_2} \cdot \nonumber\\
 &\left|\int_{-\pi/2}^{\pi/2} e^{i \cdot (3\beta+\frac{1}{2})\cdot t}
 \cdot \cos^{\beta-\frac{1}{2}}(t) \cdot
 \left(x+e^{2it}\right)^{-(\frac{1}{2}+\beta+ib\omega)}dt\right| \, .
\label{auxil1}
\end{align}
For each $t \in (-\pi/2,\pi/2)$ one now has 
\begin{equation}
x+e^{2it} = |x+e^{2it}|\cdot e^{i\cdot[t+\arctan\left(\frac{1-x}{1+x} 
\cdot \tan(t)\right)]} \, ,
\label{auxil2}
\end{equation}
and hence,
\begin{align}
 |\left(x+e^{2it}\right)^{-(\frac{1}{2}+\beta+ib\omega)}| = &
 |\left(x+e^{2it}\right)^{b \omega_2 -\frac{1}{2}-Re(\beta)} \cdot
 e^{(b \omega_1 + Im(\beta)) \cdot [t+ \arctan\left(\frac{1-x}{1+x}
 \cdot \tan(t)
 \right)]} \nonumber\\
 \leqslant &e^{\pi \cdot |Im(\beta)|} \cdot e^{2b|\omega_1| \cdot
 |t|} \cdot
 (1-x)^{-(\frac{1}{2}+Re(\beta))} \cdot |x+e^{2it}|^{b \omega_2} \\
 \leqslant &e^{\pi \cdot |Im(\beta)|} \cdot e^{2b|\omega_1| \cdot
 |t|} \cdot (1-x)^{-(\frac{1}{2}+Re(\beta))} \cdot
 \begin{cases}
 (1+x)^{b \omega_2} & \text{for $\omega_2 > 0$} \\
 (1-x)^{b \omega_2} & \text{for $\omega_2 \leqslant 0$}
 \nonumber
 \end{cases}\, .
\label{auxil3}
\end{align}
From (\ref{auxil1}), (\ref{auxil3}) follows,
\begin{align}
 |h(\omega, \beta, x)| \leqslant & \frac{\pi^{-1} \cdot
 2^{Re(\beta)-\frac{1}{2}}} {|\Gamma(\frac{1}{2}+\beta)|} \cdot
 e^{5 \pi \cdot |Im(\beta)|/2} \cdot \int_{-\pi/2}^{\pi/2}
 e^{2b|\omega_1| \cdot |t|} \cdot \cos^{Re(\beta)-\frac{1}{2}}(t)dt
 \cdot
 \nonumber\\
 & (1-x)^{-(\frac{1}{2} + Re(\beta))} \cdot
 \begin{cases}
 \left(\frac{1+x}{1-x}\right)^{b\omega_2} & \text{for
 $\omega_2 > 0 $}
 \\
 1 & \text{for $\omega_2 \leqslant 0$}
 \end{cases} 
 \nonumber\\
 = & \frac{\pi^{-1} \cdot 2^{Re(\beta)+\frac{1}{2}}}
 {|\Gamma(\frac{1}{2}+\beta)|} \cdot e^{5 \pi \cdot
 |Im(\beta)|/2} \cdot \int_{0}^{\pi/2} e^{-2b|\omega_1|
 \cdot |t|} \cdot \sin^{Re(\beta)-\frac{1}{2}}(t)dt \cdot
 \\
 & (1-x)^{-(\frac{1}{2} + Re(\beta))} \cdot
 \begin{cases}
 \left(\frac{1+x}{1-x}\right)^{b\omega_2} & \text{for
 $\omega_2 > 0 $}
 \\
 1 & \text{for $\omega_2 \leqslant 0$} \nonumber
 \end{cases} \, ,
\label{auxil4}
\end{align}
and from this (\ref{estimate1h}) by using (\ref{intest}). $_\blacksquare$

From (\ref{representation_of_K}), 
(\ref{estimate1h}) and the continuity of $K$ a straightforward calculation
provides the following estimate. 

\noindent
 {\it Lemma 8}. Let $\omega \in 
{\Bbb{C}} \setminus \left(q(A) \cup -q(A) \right)$, $x \in {\Bbb{R}}$ and $y \in {\Bbb{R}}$ 
be given. Then,
\begin{align}
 |K(\omega,x,y)| \leqslant & \pi^2 bC'_\alpha C'_{-\alpha} \cdot
 \frac{e^{2\pi b |\omega_1|}} {|\cos(2 \pi \alpha) + \cosh(2 \pi b
 \omega)|} \cdot (1+4b^2 \omega_1 ^2)^{-\frac{1}{2}} \cdot H_1(x
 ,y) \cdot
 \nonumber\\
 &
 \begin{cases}
 e^{\omega_2 \cdot D(x,y)} & \text{for $\omega_2 > 0$} \\
 e^{\omega_2 \cdot |x-y|} & \text{for $\omega_2 \leqslant 0$}
 \end{cases} \, ,
\end{align}
where,
\begin{equation}
 H_1(x,y) :=
 \begin{cases}
 (1+e^{-2x/b})^{\frac{1}{2}+Re(\alpha)} \cdot
 (1+e^{2y/b})^{\frac{1}{2}-Re(\alpha)} & \text{for $y \leqslant x$} \\
 (1+e^{-2y/b})^{\frac{1}{2}+Re(\alpha)} \cdot
 (1+e^{2x/b})^{\frac{1}{2}-Re(\alpha)} & \text{for $y > x$}
 \end{cases} \, ,
 \label{definition_of_H1}
\end{equation}
and,
\begin{equation}
 D(x,y) := |x-y|+ b \cdot
 \begin{cases}
 \ln (1+2e^{-2x/b})+\ln(1+2e^{2y/b}) & \text{for $y \leqslant x$} \\
 \ln(1+2e^{-2y/b}) + \ln(1+2e^{2x/b}) & \text{for $y > x$}
 \end{cases}.
 \label{definition_of_D}
\end{equation}
Note that the functions $H_1$ and $D$ are symmetric.
Obviously (\ref{estimate1h}) implies (\ref{estimate_2}), where,
\begin{equation}
 C_1(g,f) := \pi^2 bC'_\alpha C'_{-\alpha} \cdot
 \iint \limits_{{\Bbb{R}}^2} g^* (x)H_1(x,y)f(y)dxdy \, .
 \label{definition_of_C1}
\end{equation} 
In the following an estimate is given on the members of the 
sequence (\ref{sequence}). The derivation of this estimate uses the 
following Lemmata.

\noindent 
{\it Lemma 9}. Let $\beta \in {\Bbb{C}}$ with $ -1/2 < Re(\beta) <1/2$, 
$z \in {\Bbb{C}}$ with $|z| < 1$ and $k \in {\Bbb{N}}$ . Then the following 
recursion holds, 
\begin{align}
 & F\left( -(k+2),-(k+2)+2\beta,\frac{1}{2} +\beta-(k+2),z\right) =
 \nonumber\\ & (1-2z)F\left( -(k+1),-(k+1)+2\beta,\frac{1}{2}
 +\beta-(k+1),z\right) + \\ &
 \frac{(k+1)(k+1-2\beta)}{\left[k+2-\left(\frac{1}{2}+\beta\right)\right]
 \left[k+1-\left(\frac{1}{2}+\beta\right)\right]}z(1-z) F\left(
 -k,-k+2\beta,\frac{1}{2} +\beta-k,z\right) \, . \nonumber
\end{align} 

\noindent 
{\it Proof}. The proof is an straightforward consequence of 
Formulae~15.2.2, 15.5.1 and 5.5.3 of \cite{abramowitz}.
$_\blacksquare$
 
\noindent
{\it Lemma 10}. Let $\beta \in {\Bbb{C}}$ with $ -1/2 < Re(\beta) <1/2$ 
and $y \in [0,1)$ Then for every $k \in {\Bbb{N}}$ 
the following estimate holds,
\begin{equation} \label{estlemma10}
\left|F\left( -k,-k+2\beta,\frac{1}{2} +\beta-k,y\right)\right| 
\leqslant 1 \, .
\end{equation}

\noindent
{\it Proof}. The estimate (\ref{estlemma10}) follows from Lemma 9 using 
induction along with the following estimate for each $k \in {\Bbb{N}}$, 
\begin{equation} \label{est2}
\left| \frac{(k+1)(k+1-2\beta)}{\left[k+2-\left(\frac{1}{2}+\beta\right)
\right]\left[k+1-\left(\frac{1}{2}+\beta\right)\right]} \right| 
\leqslant 2 \, .
\end{equation}
$_\blacksquare$

The following inequalities for the elements of the quasinormal mode 
sequence (\ref{sequence}) are straightforward consequences of the 
definitions and Lemma~10 (as well as of Formulae~6.1.17, 6.1.26, 
6.1.22 of \cite{abramowitz}). For given $x \in {\Bbb{R}}$, 
$y\in {\Bbb{R}}$, $t\in {\Bbb{R}}$ and $k \in {\Bbb{N}}$ one gets,
\begin{equation} \label{estqnms}
\left| c_{\omega_k^{\pm}}u_{\omega_k^{\pm}}(y) 
u_{\omega_k^{\pm}}(x)e^{i{\omega_k^{\pm}}t} \right| \leqslant
a_k^{\pm}
\left(e^{-\left(t-D_s(x,y)\right)/b}\right)^
{\frac{1}{2}\pm Re(\alpha)+k} \, ,
\end{equation}
where 
\begin{equation} \label{Koeffiz}
a_k^{\pm} := 
\frac{1}{4\pi}
\left|\cot(\pi \alpha)\right|
\frac{\Gamma(1 \pm 2 Re(\alpha))}{\left|\Gamma(1 \pm 2 \alpha)\right|}
\frac{\left(\Gamma\left(\frac{1}{2}\pm Re(\alpha)+k\right)\right)^2}
 {\Gamma(1+k) \Gamma(1 \pm 2 Re(\alpha)+k)} \, .
\end{equation}
Further using 6.1.22 of \cite{abramowitz} it is easy to see that there is
a positive constant $C_{\alpha}$ such that 
\begin{equation} \label{estKoeffiz}
\left|a_k^{\pm}\right| \leqslant C_{\alpha} \, .
\end{equation}
Hence with such a constant $C_{\alpha}$ one gets for given 
$x \in {\Bbb{R}}$, $y\in {\Bbb{R}}$, 
$t\in {\Bbb{R}}$ and $k \in {\Bbb{N}}:$
\begin{equation} \label{festqnms}
\left| c_{\omega_k^{\pm}}u_{\omega_k^{\pm}}(y) 
u_{\omega_k^{\pm}}(x)e^{i{\omega_k^{\pm}}t} \right| \leqslant
C_{\alpha}
\left(e^{-\left(t-D_s(x,y)\right)/b}\right)^
{\frac{1}{2}\pm Re(\alpha)+k} \, .
\end{equation}
For given $x \in {\Bbb{R}}$ and $y\in {\Bbb{R}}$ from the last 
estimate follows the absolute and uniform summability of 
\begin{equation} \label{Folge}
\left( c_{\omega} u_{\omega}(y) u_{\omega}(x) e^{i \omega t} 
\right)_{\omega \in q(A)}
\, ,
\end{equation}
on every compact subset of $(D_s(x,y),\infty) \times {\Bbb{R}}$ and 
hence also the analyticity of the function which associates to each 
$t \in (D_s(x,y),\infty) \times {\Bbb{R}}$ the value
\begin{equation} 
\sum_{\omega\in q(A)}c_\omega u_\omega(y)u_\omega(x)e^{i \omega t}\, .
\label{Greensfkt}
\end{equation}
A further consequence of the estimate is that 
 \begin{equation}
 \left( c_\omega \cdot \int\limits_{-\infty}^{+\infty} u_\omega (y)f(y)dy
 \cdot \int\limits_{-\infty}^{+\infty} g^*(x)u_\omega (x)dx 
 \cdot e^{i \omega t}\right)_{\omega \in q(A)} \, ,
\end{equation}
is {\it absolutely and uniformly summable} on $[t_0,\infty) \times K_0$, 
where $t_0 > M_s(g,f)$, $K_0$ is any compact subset of ${\Bbb{R}}$ and
\begin{equation} 
M_s(g,f) := \max \{ D_s(x,y): x \in \text{supp($g$)} \;\; 
\text{and} \;\; y \in \text{supp($f$)} \} \, .
\end{equation}
Hence also the analyticity of the function which associates to each 
$t \in (M_s(x,y),\infty) \times {\Bbb{R}}$ the value
\begin{equation}
\sum_{\omega \in q(A)}c_\omega \cdot \int\limits_{-\infty}^{+\infty} 
u_\omega (y)f(y)dy\cdot \int\limits_{-\infty}^{+\infty} 
g^*(x)u_\omega (x)dx \cdot e^{i \omega t} \, .
\end{equation}
The remainder of this appendix considers the sequence 
\begin{equation} 
\label{nonsummable}
\left(c_{\omega}\left[u_{\omega}(0)\right]^2\right)_{\omega\in q(A)}
\, , 
\end{equation}
which is a special case of {(\ref{Folge})} for $x=y=0$ and $t=0$.
This is interesting because for this case $t<D_s(x,y)$, which was not
considered up to now. In the following it will be shown that {\it this
sequence is not absolutely summable}.
 
First, after some computation, which uses Formulae~15.4.19, 8.6.1,
6.1.17, 6.1.18, of \cite{abramowitz}, it can be seen that
\begin{equation}
u_{\omega_{2k+1}^{-}}(0) = 0 \, ,
\end{equation}
and that 
\begin{equation} \label{divergent}
\left| c_{\omega_{2k}^{-}} \left[u_{\omega_{2k}^{-}}(0)\right]^2 \right|=
\frac{1}{2\pi} |\cot(\pi\alpha)| \frac{\Gamma(k+\frac{1}{2}) 
|\Gamma(k-\alpha+\frac{1}{2})|}{\Gamma(k+1) |\Gamma(k-\alpha+1)|} \, .
\end{equation}
both for each $k \in {\Bbb{N}}$. Further, using Formulae~6.1.17, 6.1.26, 
6.2.1 of \cite{abramowitz} (as well as Fubini's theorem and Tonelli's 
theorem) one gets for each $n \in {\Bbb{N}}$,
\begin{align} \label{integralestimate}
 & \sum_{k=0}^n \left| c_{\omega_{2k}^{-}}
 \left[u_{\omega_{2k}^{-}}(0)\right]^2 \right| \geqslant \frac{1}{2
 \pi} \frac{|\cos(\pi Re(\alpha))|}{|\sin(\pi\alpha)|} \sum_{k=0}^n
 \frac{\Gamma(k+\frac{1}{2}) \Gamma(k- Re(\alpha)+\frac{1}{2})}
 {\Gamma(k+1) \Gamma(k- Re(\alpha)+1)} \nonumber\\ &=\frac{1}{2 \pi^2}
 \frac{|\cos(\pi Re(\alpha))|}{|\sin(\pi\alpha)|} \iint
 \limits_{(0,1)^2} \frac{1-(ts)^{n+1}}{1-ts}
 \left[ts(1-t)(1-s)\right]^{-1/2}s^{-Re(\alpha)}dtds \, .
\end{align}

The proof that (\ref{nonsummable}) is not absolutely summable proceeds 
indirectly. From the assumption that it is absolutely summable it 
follows by (\ref{integralestimate}) and the monotonous convergence 
theorem that the function defined by 
\begin{equation}
 (1-ts)^{-1} \left[ts(1-t)(1-s)\right]^{-1/2}s^{-Re(\alpha)} \, ,
\end{equation}
for each $t\in (0,1)$ and $s\in (0,1)$ is integrable on $(0,1)^2$ . 
Hence using the substitution
\begin{equation}
t= \sin^2(\tau) \quad , \quad \tau \in (0, \pi/2) \, ,
\end{equation}
and Fubini's theorem it follows that the function defined by
\begin{equation}
s^{-(\frac{1}{2}+Re(\alpha))} (1-s)^{-1} \, ,
\end{equation}
for each $s\in (0,1)$ is integrable on $(0,1)$, which is false.
Hence (\ref{nonsummable}) is not absolutely summable.

Using similar methods it can be shown that the sequence
\begin{equation}
\sum_{k=0}^n 
\left(c_{\omega_k^{-}} \left[u_{\omega_k^{-}}(0)\right]^2 +
c_{\omega_k^{+}} \left[u_{\omega_k^{+}}(0)\right]^2  
\right) \, ,
\end{equation}
converges for $n \to \infty$. The proof of this is not given here.
Note that this does not contradict the fact that (\ref{nonsummable})
is not absolutely summable. The ``sum'' of (\ref{nonsummable}) just
{\it depends on the order of the summation}.

\section{ }

\label{appendix2}
This appendix gives a (not completely rigorous) derivation for the
representation (\ref{representation_of_phi_g}) used in this paper for
the solution to the initial value problem. The derivation uses the
``Laplace method'' of \cite{schmidt}. To aid further applications of
this general method, the derivation is provided for a more general
wave equation (\ref{wave_equation_4}) than used in this paper.

Take as given $J$, $V$, and $f$, where $J$ is a non empty (bounded or
unbounded) interval of ${\Bbb{R}}$, $V$ is a continuous real-valued
function on $J$, and $f$ is a square integrable complex-valued
function on $J$. Let $\phi_f$ be a given complex-valued function on
${\Bbb{R}} \times J $, which is two times continuously partially
differentiable and which satisfies
\begin{equation}
\frac{\partial^2 \phi_f}{\partial t^2}(t,x) 
-\frac{\partial^2 \phi_f}{\partial x^2}(t,x)
+V(x)\phi_f(t,x) = 0 \, , 
\label{wave_equation_4}
\end{equation}
for each $t \in \Bbb{R}$ and $x \in J$.
In addition $\phi_f$ satisfies the initial conditions
\begin{equation}
\phi_f(0,x) = 0, 
\quad
\frac{\partial \phi_f}{\partial t}(0,x) = f(x) \, ,
\label{initial_conditions_4}
\end{equation}
for each $t \in \Bbb{R}$ and $x \in J$. Finally, let $\epsilon$ 
be a given strictly positive real (otherwise arbitrary) number.

By Laplace transforming (\ref{wave_equation_4}) and using 
(\ref{initial_conditions_4}) one gets the representations 
(\ref{equation_for_calF}), (\ref{representation_of_phi}) of $\phi_f$
below as follows. Defining 
\begin{equation}
\psi_f(t,x) := e^{- \epsilon t} \phi_f (t,x) \, ,
\label{definition_of_psi}
\end{equation}
for each $t \in {\Bbb{R}}$, $x \in J$ and assuming the boundedness of 
\begin{equation}
\psi_f(\cdot , x) \, , 
\quad
\frac{\partial \psi_f }{\partial t}(\cdot , x) \, ,
\quad
\frac{\partial^2 \psi_f }{\partial t^2}(\cdot , x) \, ,
\label{conditions_1}
\end{equation}
on each $[0, \infty)$ for each $x \in J$ one gets from 
(\ref{wave_equation_4}), (\ref{initial_conditions_4}) for arbitrary 
$x \in J$ and $s \in (0,\infty) \times {\Bbb{R}}$,
\begin{equation}
\int_0^\infty e^{-st} \cdot 
\left(-\frac{\partial^2 \psi_f}{\partial x^2}(t,x) 
+ [V(x) + (s+ \epsilon )^2 ]
\cdot \psi_f(t,x) \right) dt = f(x) \, .
\label{Laplace_transform_of_wave_equation_4} 
\end{equation}
From this, assuming the uniformly boundedness of 
\begin{equation}
\psi_f(\cdot , y) , \quad
\frac{\partial \psi_f }{\partial t}(\cdot , y), \quad
\frac{\partial^2 \psi_f }{\partial t^2}(\cdot , y) \, ,
\label{conditions_2}
\end{equation}
for $y$ from a neighbourhood of $x$, one concludes
\begin{equation}
- (\Psi_f(s,\cdot))''(x) + [V(x)-(-is-i\epsilon)^2 ]
\Psi_f(s,x) = f(x) \, ,
\label{equation_for_Psi}
\end{equation}
where 
\begin{equation}
\Psi_f(s,y) := \int_0^\infty e^{-st}\psi_f(t,y)dt \, ,
\label{definition_of_Psi}
\end{equation}
for each $y \in J$. Note that, roughly speaking, the $-i\epsilon$ 
also guarantees the unique solvability of (\ref{equation_for_Psi}) in 
$L^2({\Bbb{R}})$ for the limiting cases where $s$ is purely imaginary. 
This fact will be used in (\ref{definition_of_Psi}) for inverting the 
Laplace transform. Formal inversion of (\ref{equation_for_Psi}) leads to 
\begin{align}
\Psi_f(s,x) & = G_f((\omega-i\cdot (\sigma + \epsilon) )^2, x)
\nonumber\\
& := \int_J G( (\omega - i(\sigma + \epsilon) )^2 , x , y)f(y)dy \, ,
\label{representation_of_Psi}
\end{align}
where $G(\omega-i(\sigma+\epsilon) )^2,\cdot,\cdot)$ is a Green's 
function for the formal differential operator
\begin{equation}
-\frac{d^2}{dx^2}+V-[\omega-i(\sigma+\epsilon) ]^2 \, , 
\label{formal_operator}
\end{equation}
which one arrives at by the method of variation of constants. 
Here $\sigma$, $\omega$ denote the real and imaginary parts of $s$, 
respectively. Note that for the choice of an appropriate Green's function
it may be necessary to impose further boundary conditions on the solutions 
of (\ref{wave_equation_4}). By assuming the square integrability of 
$\psi_f(\cdot,x)$ on $[0,\infty)$ the inversion of the Laplace transform 
in (\ref{definition_of_Psi}) can be performed using the Fourier 
inversion theorem for square integrable functions on the real line. 
In this way one gets from (\ref{definition_of_Psi}), 
(\ref{representation_of_Psi}) the representations,
\begin{equation}
F^{-1} G_f((\cdot-i\epsilon)^2,x) = \sqrt{2\pi} \cdot 
\begin{cases}
\psi_f(t,x) & \text{for $t \geqslant 0$} \\
0 & \text{for $t < 0$}
\end{cases} \, ,
\label{equation_for_calF}
\end{equation}
where F denotes the unitary linear Fourier transformation on 
 $L^2(\Bbb{R})$ 
(defined according to \cite{reedsimon}, Volume~II) as well as for 
(Lebesgue-) almost all $t \in [0,\infty)$,
\begin{equation}
\phi_f(t,x)=\frac{1}{2\pi}\lim_{\nu\to\infty}\int \limits_{-\nu}^{\nu} 
e^{it\cdot(\omega-i\epsilon)}G_f((\omega-i\epsilon)^2,x)d\omega\, .
\label{representation_of_phi}
\end{equation}
Note that the limit in the last formula is essential since from 
the assumptions made one can only conclude that the integrand is 
square integrable (but not integrable) over ${\Bbb{R}}$. 
Moreover note that the right hand side of (\ref{representation_of_Psi}) 
and the left hand side of (\ref{representation_of_phi}) are independent 
of $\epsilon$, which reflects the fact that in the inversion of the 
Laplace transformation there is some freedom in the choice of contour.
Starting from (\ref{representation_of_Psi}) one can arrive at 
(\ref{representation_of_phi_g}) in the following way. Let $g$ be an 
arbitrarily chosen infinitely often differentiable complex-valued 
function on ${\Bbb{R}}$ having compact support. Assuming the uniform 
boundedness of $\psi_f$ on ${\Bbb{R}} \times \operatorname{supp}(g)$ 
one gets from (\ref{definition_of_Psi}), (\ref{representation_of_Psi}),
\begin{equation}
\Psi _{g,f} (s,x) = G_{g,f}((\omega -i \cdot \epsilon)^2) \, ,
\label{representation_of_Psig}
\end{equation}
where
\begin{align}
 \Psi_{g,f} (s) & := \int_{0}^{\infty} e^{-st} \cdot <g|\psi_f
 (t,\cdot)>dt \, , 
 \nonumber\\
 <g|\psi_f(t,\cdot)>&:=\int_{J}^{} g^* (x) \psi_f (t,x) dx \quad
 \text{for each $ t\in \Bbb{R}$} \, ,
 \label{definitions_of_Psigusw}
 \\
 G_{g,f}((\omega -i \cdot \epsilon)^2) & := \int_J g^* (x) G_f
 ((\omega -i \cdot \epsilon)^2,x)dx \, . \nonumber
\end{align}
Assuming the square integrability of the function which associates to
each $t \in [0,\infty)$ the value of $<g|\psi_f(t,\cdot)>$ one gets
from (\ref{representation_of_Psig}) by the Fourier inversion theorem,
\begin{equation}
[F^{-1} G_{g,f}((\cdot-i\epsilon)^2)](t) = \sqrt{2\pi} \cdot 
 \begin{cases}
 <g|\psi_f(t,\cdot)> & \text{for $t \geqslant 0$} \\
 0 & \text{for $t < 0$}
 \end{cases} \, ,
 \label{equation_for_calFg}
\end{equation}
as well as for almost all $t \in [0,\infty)$,
\begin{equation}
 <g|\phi_f(t,\cdot)> = \frac{1}{2\pi} \lim_{\nu \to \infty}
 \int \limits_{-\nu}^{\nu} 
 e^{it \cdot (\omega-i\epsilon)} G_{g,f}((\omega - i\epsilon)^2)
 d\omega \, .
 \label{representation_of_gphi}
\end{equation}
Formula (\ref{representation_of_gphi}) is easily seen to be a
consequence of (\ref{equation_for_calFg}).  In the following,
sufficient conditions are given for the validness of the
Formulae~(\ref{equation_for_calFg}) and
(\ref{representation_of_gphi}) leading to the
formulae~(\ref{representation_of_phieta}) and
(\ref{representation_of_phietat}), respectively. For the terminology
used in the following theorem consult for example Volume~I of
\cite{reedsimon}.
 
\noindent
{\it Theorem 1}. 
Let $X$ be a non trivial complex Hilbert space with the scalar 
product $<|>$. Let $A:D(A) \rightarrow X$ be a densely defined, 
linear, self-adjoint, semibounded operator in X with spectrum 
$\sigma (A)$ and resolvent $R$, where the latter is defined by 
$R(\lambda) := (A-\lambda)^{-1}$ for each $\lambda \in {\Bbb{C}} 
\setminus \sigma (A)$. Define
\begin{equation}
 \alpha := \begin{cases}
 0 & \text{for $\min \sigma(A) \geqslant 0$} \\ 
 \sqrt{-\min \sigma (A)} & \text{for $\min \sigma(A) < 0$} 
 \end{cases} \, .
 \label{definition_of_alpha}
\end{equation}
Also for each $\xi , \eta \in X$ define the analytic function 
${\cal R}_{\xi,\eta} : {\Bbb{R}} \times (-\infty, -\alpha) \rightarrow 
\Bbb{C}$ by 
\begin{equation}
{\cal R}_{\xi,\eta} (\omega+i\sigma) := 
<\xi|R((\omega+i\sigma)^2)\eta> \, ,
\label{definition_of_calRxieta}
\end{equation}
for each $\omega \in {\Bbb{R}}$ and $\sigma \in (-\infty,-\alpha)$.
Finally, let $\xi$ and $\eta$ be arbitrary elements of $D(A)$ and 
$X$, respectively and let $\phi_{\xi}$ be the unique element of 
$C^2({\Bbb{R}},X)$ satisfying for each $t \in {\Bbb{R}}$
\begin{equation}
\phi_{\xi} '' (t) = -A\phi_{\xi}(t) \, ,
\label{wave_equation_5}
\end{equation}
and the initial conditions 
\begin{equation}
\phi_{\xi}(0) = 0 , \qquad \phi_{\xi}' (0) = \xi \, .
\label{initial_conditions_5}
\end{equation}
Then for each $\epsilon \in (\alpha, \infty)$ and almost all (in 
the Lebesgue sense) $t \in \ [0, \infty)$,
\begin{equation} \label{representation_of_phieta}
<\eta|\phi_{\xi}(t)> = \frac{e^{\epsilon t}}{\sqrt{2\pi}}
[F^{-1}{\cal R}_{\eta,\xi}(\cdot-i\epsilon)](t) \, ,
\end{equation}
and
\begin{equation}
<\eta|\phi_{\xi}(t)> = \frac{1}{2\pi}
\lim_{\nu \to \infty}
\int \limits_{-\nu}^{\nu} 
e^{it \cdot (\omega-i\epsilon)} {\cal R}_{\eta,\xi} (\omega - i\epsilon)
d\omega \, .
\label{representation_of_phietat}
\end{equation}
This theorem is mainly a consequence of Theorem~11.6.1 in 
\cite{hillephillips} and the proposition on Page~295 in Volume~II 
of \cite{reedsimon}, and will not be proved here.

\newpage

\newpage


\begin{thebibliography}{99}
 
\bibitem{abramowitz} Abramowitz M and Stegun I A (ed) 1984 {\em
 Pocketbook of Mathematical Functions} (Thun: Harri Deutsch)
 
\bibitem {motetbachelot} Bachelot A and Motet-Bachelot A 199? Les
 resonances d'un Trou Noir de Schwarzschild {\em Ann. I.H.P. physique
 theorique} {\bf ?} ?
 
\bibitem{beyer94} Beyer H R 1995 The spectrum of radial adiabatic
 stellar oscillations {\em J. Math. Phys.} {\bf 36} 4815-4825
 
\bibitem{chandra} Chandrasekhar S and Detweiler S 1975 The
 quasi-normal modes of the Schwarzschild black hole {\em Proc. R.
 Soc. Lond. A.} {\bf 344} 441-452
 
\bibitem{ferrarimashoon} Ferrari V and Mashoon B 1984 New approach to
 the quasinormal modes of a black hole {\em Phys. Rev. D} {\bf 30}
 295-304
 
\bibitem{hillephillips} Hille E and Phillips R S 1957 {\em Functional
 Analysis and Semi-Groups} (Providence: AMS)
 
\bibitem{oberhettinger} Oberhettinger F 1973 {\em Fourier Transforms
 of Distributions and their Inverses} (New York: McGraw-Hill)
 
\bibitem{poeschlteller} P\"{o}schl G. and Teller E 1933 Bemerkungen
 zur Quantenmechanik des harmonischen Oszillators {\em Z. Phys.} {\bf
 83} 143-151
 
\bibitem{pricehusain} Price R H and Husain V 1992 Model for the
 completeness of quasinormal modes of relativistic stellar
 oscillations {\em Phys. Rev. Lett.} {\bf 68} 1973-1976
 
\bibitem{reedsimon} Reed M and Simon B 1980, 1975, 1979, 1978 {\em
 Methods of Mathematical Physics Volume I, II, III, IV} (New York:
 Academic)
 
\bibitem{schmidt} Schmidt B G and Nollert H-P 1992 Quasinormal modes
 of Schwarzschild black holes: Defined and calculated via Laplace
 transformation {\em Phys. Rev. D} {\bf 45} 2617-2627
 
\bibitem{vishveshwara} Vishveshwara C V 1970 Stability of the
 Schwarzschild Metric {\em Phys. Rev. D} {\bf 1} 2870-2879
 
\bibitem{weidmann} Weidmann J 1976 {\em Lineare Operatoren in
 Hilbertraeumen} (Teubner: Stuttgart)
 
\bibitem{yosida} Yosida K 1980 {\em Functional Analysis} (Berlin:
 Springer)

\end{thebibliography}
\end{document}